# A forward modelling approach to AGN variability – method description and early applications

Lia F. Sartori,[1,2] Benny Trakhtenbrot,[3] Kevin Schawinski,[1,4] Neven Caplar,[5] Ezequiel Treister,[2] and Ce Zhang[6]

[1]*Institute for Particle Physics and Astrophysics, ETH Zürich, Wolfgang-Pauli-Str. 27, CH-8093 Zürich, Switzerland*
[2]*Instituto de Astrofísica, Facultad de Física, Pontificia Universidad Católica de Chile, Casilla 306, Santiago 22, Chile*
[3]*School of Physics and Astronomy, Tel Aviv University, Tel Aviv 69978, Israel*
[4]*Modulos AG, Technoparkstrasse 1, CH-8005 Zürich, Switzerland*
[5]*Department of Astrophysical Sciences, Princeton University, 4 Ivy Ln., Princeton, NJ 08544, USA*
[6]*Systems Group, ETH Zürich, Universitätstrasse 6, CH-8006 Zürich, Switzerland*

## ABSTRACT

We present a numerical framework for the variability of active galactic nuclei (AGN), which links the variability of AGN over a broad range of timescales and luminosities to the observed properties of the AGN population as a whole, and particularly the Eddington ratio distribution function (ERDF). We have implemented our framework on GPU architecture, relying on previously published time series generating algorithms. After extensive tests that characterise several intrinsic and numerical aspects of the simulations, we describe some applications used for current and future time domain surveys and for the study of extremely variable sources (e.g., "changing look" or flaring AGN). Specifically, we define a simulation setup which reproduces the AGN variability observed in the PTF/iPTF survey, and use it to forward model longer light curves of the kind that may be observed within the LSST main survey. Thanks to our efficient implementations, these simulations are able to cover for example over 1 Myr with a roughly weekly cadence. We envision that this framework will become highly valuable to prepare for, and best exploit, data from upcoming time domain surveys, such as for example LSST.

*Keywords:* AGN variability - quasars

## 1. INTRODUCTION

Variable emission is a ubiquitous property of active galactic nuclei (AGN). It can be observed or inferred at essentially all timescales, from hours to billions of years (e.g. Goyal et al. 2013; McHardy et al. 2004; Smith et al. 2018 MacLeod et al. 2012; Sartori et al. 2018b; Schawinski et al. 2015; Novak et al. 2011), and across the entire electromagnetic spectrum (e.g. Uttley & Mchardy 2004; Paolillo et al. 2004; Paolillo et al. 2017; Assef et al. 2018; Caplar et al. 2017). Since the AGN luminosity originates from the accretion process, the study of AGN variability can provide crucial information about the physics of the supermassive black hole (SMBH) and central engine, including at spatial scales that are beyond the resolving power of most of the current and future facilities for most of the AGN. In addition, it allows to probe the close link between SMBHs and their hosts by taking into account the fact that the energy injection from the AGN varies with time (e.g. Hickox et al. 2014). AGN variability is also often employed as an efficient method to select AGN from large multi-epoch surveys (e.g., Trevese et al. 2008; Villforth et al. 2010; De Cicco et al. 2015; Sánchez-Sáez et al. 2019).

In the past decades, different campaigns with recurrent photometry and/or spectroscopy such as the Sloan Digital Sky Survey (SDSS) Stripe 82 (York et al. 2000; Ivezić et al. 2007; Sesar et al. 2007), the (intermediate) Palomar Transient Factory (PTF/iPTF, Bellm 2014), the Palomar Observatory Sky Survey (POSS, Minkowski & Abell 1963) and the Catalina Real-Time Transient Survey (CRTS, Drake et al. 2009; Djorgovski et al. 2011) have provided important insights into the variability behaviour of the AGN population on days to decades timescales (e.g. MacLeod et al. 2012; Morganson et al.

Corresponding author: Neven Caplar
ncaplar@princeton.edu

2014; Caplar et al. 2017; Graham et al. 2017). Indeed, these works show that the amplitude of variability is increasing for increasing timescales, and investigate additional dependencies with physical parameters of the AGN. For a fixed timescale, the amplitude of variability is observed to anti correlate with luminosity, rest frame wavelength and Eddington ratio, while little or no dependence is found as a function of redshift (e.g. Wilhite et al. 2008; Ai et al. 2010; MacLeod et al. 2010; Caplar et al. 2017; Rumbaugh et al. 2018). The dependence on black hole mass is still unclear since different studies found either positive, negative or absent correlation (e.g. Wilhite et al. 2008; Kelly et al. 2009; MacLeod et al. 2010; Zuo et al. 2012; Caplar et al. 2017). Interestingly, the derived dependencies, as well the observed timescales at which variability is occurring in the optical and UV, cannot be fully explained with standard accretion disc theory (e.g. MacLeod et al. 2010; Zuo et al. 2012; Caplar et al. 2017).

In addition to these overall AGN variability properties, in the last years new intriguing objects such as the changing look quasars (or more generally changing look AGN, CL-AGN) have been found in increasingly large numbers (e.g. LaMassa et al. 2015; Runnoe et al. 2016; Ruan et al. 2016; McElroy et al. 2016; Yang et al. 2018; Mathur et al. 2018; Wang et al. 2018; Stern et al. 2018; Ross et al. 2018; Zetzl et al. 2018; Katebi et al. 2018; MacLeod et al. 2019; Trakhtenbrot et al. 2019). CL-AGN are characterised by the appearance or disappearance of broad Balmer emission line components over periods ranging from months to years[1], often accompanied by changes in luminosity of over one order of magnitude in the same period, which again challenges our understanding of AGN accretion physics. Indeed, the observed large magnitude changes in short timescales are not consistent with the predictions from standard thin accretion disc (e.g. Krolik 1999; LaMassa et al. 2015; MacLeod et al. 2016). CL-AGN also challenge the standard AGN unification model (Antonucci 1993; Urry & Padovani 1995), where the presence and absence of broad emission lines is explained by different line of sight orientations, as the angular position of the obscuring material is not expected to vary on such short timescales. One open question is if a similar level of variability should be expected for every AGN, or if different AGN may belong to distinct classes with significantly different variability properties and physical mechanisms.

As mentioned above, the rapid variability observed in the optical/UV continuum, in particular in extreme cases such as the CL-AGN, is hard to reconcile with standard accretion disc theory. Indeed, months to years timescales are much shorter compared to the viscous timescales, at which variability due to overall accretion changes would appear, which is on the order of $\sim 100 - 1000$ yr (see Lawrence 2018 for a discussion of the "viscosity crisis"). The observed timescales are more consistent with thermal timescales, which are relevant for instabilities or local perturbations in the accretion disc. Such instabilities and perturbations could explain the moderate variability characteristic of the majority of the AGN, as well as produce structural changes in the inner part of the accretion disc which could lead to the extreme variability observed in CL-AGN (e.g. Stern et al. 2018; Ross et al. 2018). The possible causes of such instabilities and perturbations are still matter of debate, but they include magnetorotational instabilities (MRI, Balbus & Hawley 1991; Reynolds & Miller 2009), local temperature fluctuations driven e.g. by X-ray heating (Shappee et al. 2013), iron opacity (Jiang et al. 2016) as well as perturbations due to stellar mass black holes, stellar remnants and stars moving within the dense medium of the accretion disc (e.g. Syer et al. 1991; McKernan et al. 2014; Bartos et al. 2017). Other possible explanations are that accretion discs are magnetically elevated, which could lead to larger scale heights and shorter variability timescales compared to what expected for standard thin accretion discs (Dexter & Begelman 2019), or that the behaviour of the accretion disc is dominated by non-local physics such as magnetic fields connecting different regions on timescales shorter than viscous timescales (Lawrence 2018).

The above refers to variability which can be directly traced through observations. In addition to this, indirect arguments based on the photoionization state of large scale gas in (and outside) galaxies indicate that AGN may dramatically change their luminosity on timescales which are significantly longer than what we can directly probe, i.e. $\gtrsim 10^4$ yr (e.g. Lintott et al. 2009; Keel et al. 2012a; Gagne et al. 2014; Schawinski et al. 2015; Sartori et al. 2016; Sartori et al. 2018b). This type of long timescale variability is addressed by simulations (e.g. Novak et al. 2011; Gabor & Bournaud 2013) and theoretical models (e.g. Martini & Schneider 2003; King & Nixon 2015), and may have implications for our attempts to understand the SMBH-host galaxy co-evolution (Hickox et al. 2014; Volonteri et al. 2015).

---

[1] Another group of so called CL-AGN are AGN whose X-ray spectra show a switch from Compton thick to Compton thin, or vice versa (e.g. Matt et al. 2003; Piconcelli et al. 2007; Marchese et al. 2012; Ricci et al. 2016). The relation between "optical CL-AGN" and "X-ray CL-AGN" is not clear yet.

With the advance of large, multiwavelength time domain surveys such as the Large Synoptic Survey Telescope (LSST, Ivezic et al. 2008; LSST Science Collaboration et al. 2009), the Time-Domain Spectroscopic Survey in SDSS-IV (TDSS, Morganson et al. 2015), SDSS-V (Kollmeier et al. 2017), the Zwicky Transient Facility (ZTF, Bellm et al. 2019), the All-Sky Automated Survey for SuperNovae (ASAS-SN, Shappee et al. 2014) and the extended ROentgen Survey with an Imaging Telescope Array (eROSITA, Merloni et al. 2012) we are entering an exciting era of time domain astronomy which will allow us to probe the variable Universe with an unprecedented cadence (∼days), depth, sky area, and time-span. These surveys will allow us to probe the variability properties of individual AGN, of the AGN population as a whole, as well as new types of flares and other extreme variability phenomena associated with SMBH accretion such as CL-AGN or tidal disruption events (TDE; e.g., Rees 1988; van Velzen et al. 2011; Gezari et al. 2012; Arcavi et al. 2014; Chornock et al. 2014; Hung et al. 2017; Auchettl et al. 2017). This will provide crucial information to quantify the variability phenomena and to test various models for its origin. However, challenges for modelling and characterisation of the observed light curves will be important. These are due to the finite time resolution and length of the obtained light curves (e.g. Uttley et al. 2002; Emmanoulopoulos et al. 2010), as well as sample selection and observational biases inherent to each survey. A careful study of AGN variability therefore requires sophisticated statistical analysis methods as well as computationally expensive modelling and simulations.

In Sartori et al. (2018a) we proposed that the AGN variability observed at optical/UV wavelengths may be modelled based on the distribution of the Eddington ratio ($L/L_{\rm Edd}$) among the (observed) AGN population. In fact, the emission at such wavelengths mainly arises from the accretion process (Shakura & Sunyaev 1973) and can therefore be modulated by changes in $L/L_{\rm Edd}$, although other processes such as reprocessing of high energy photons from the hot corona may as well affect the observed luminosity at short timescales (e.g. Uttley et al. 2003). Specifically, we suggest that AGN light curves can be fully simulated starting from two statistical functions: the Eddington ratio distribution function (ERDF, representing the $L/L_{\rm Edd}$ probability density function PDF, see Section 2.1.1) defining the possible $L/L_{\rm Edd}$ values in the simulation, and a power spectral density (PSD, see Section 2.1.2) describing the variability and therefore the time ordering of the $L/L_{\rm Edd}$ points. The obtained $L/L_{\rm Edd}$ time-series can then be converted to (optical/UV) light curves in the observed band in question following commonly-used conversion factors (i.e., reciprocal bolometric corrections). In this framework, every AGN light curve is therefore one realisation of the underlying variability process described by the assumed ERDF+PSD set, and represents the entire AGN population (following the ergodic hypothesis). A forward modelling approach allows to compare the predictions of our models to observations to test if and how the observed variability features can be reproduced with this simple model and, if this is the case, constrain the underlying ERDF and PSD. In addition, given a model, it is possible to produce light curves that would be observed by specific facilities. We also stress that what proposed in Sartori et al. (2018a) allows to link and discuss, in the same framework, AGN variability observed at very different timescales.

In this paper we present a new simulations setup to produce AGN light curves following the framework proposed in Sartori et al. (2018a), based on the algorithm presented in Emmanoulopoulos et al. (2013) (E13 hereafter), and discuss possible applications. The paper is structured as follows. In Section 2 we describe the model and method. An extensive description of the simulations code and of the tests to characterise both intrinsic and numerical behaviours of the simulations is given in Section 3. Finally, in Section 4 we discuss some early application of our framework based on the observed variability in the PTF/iPTF survey and aimed at interpreting extremely variable SMBHs in the LSST era. An up-to-date version of the simulations code, along with examples and additional information about its usage, can be found at: https://github.com/nevencaplar/AGN-Variability-Simulations.

## 2. MODEL AND METHOD

The goal of our framework is to statistically model the (observed) AGN variability due to changes in accretion (thus, variability in Eddington ratio, $L/L_{\rm Edd}$), by simulating $L/L_{\rm Edd}$ time series based on the distribution of $L/L_{\rm Edd}$ among the (active) galaxy population. Specifically, we simulate $L/L_{\rm Edd}$ time series whose PDF is inspired by the ERDF, and with variability behaviour described by a chosen input PSD.

In this section we outline the methods that we use to simulate the $L/L_{\rm Edd}$ time series. Specifically, we discuss how we define the input PDF and PSD, and how the simulated $L/L_{\rm Edd}$ time series can be converted to (observed) AGN light curves. Details about the code implementation and testing are given in Section 4. Finally, some early applications aimed at illustrating the potential of our new framework are discussed in Section 3.

We note that, although the framework presented here and the algorithm are optimised for AGN and $L/L_{\rm Edd}$ time series, they can be adapted to investigate other variable processes such as, e.g. stellar variability (e.g. Catelan & Smith 2015 and references therein; Labadie-Bartz et al. 2017) or star formation histories (Caplar & Tacchella 2019).

### 2.1. *Simulations of Eddington ratio time series*

To simulate $L/L_{\rm Edd}$ time series we start from the time series generating algorithm proposed by E13. This algorithm, which is based on previous algorithms presented in Timmer & Koenig (1995) (TK95 hereafter) and Schreiber & Schmitz (1996) (S96 hereafter), produces time series whose PDF and PSD are consistent with the desired (input) ones. These input PDF and PSD can be taken from real observations (e.g. to reproduce multiple observations with the same variability behaviour as an observed light curve), or from models which have to be tested against the data. The first step consists of creating a time series of the desired length (e.g., in years) and time resolution, whose periodogram scatters around the underlying input PSD (TK95 algorithm), therefore determining the variability in the data. A second time series with the desired number of steps and PDF corresponding to the input one is then obtained through a random draw. Finally, the two time series are combined following the iterative amplitude adjusted Fourier transform algorithm described S96 to produce a time series with both PDF and PSD consistent with the input ones (see Fig. 1 in E13). We stress that although this algorithm can be used to generate the time evolution of any given quantity (e.g., count rates, fluxes, or magnitudes), in our framework it is only employed to directly simulate $L/L_{\rm Edd}$ time series, while the conversion from $L/L_{\rm Edd}$ to other observed quantities is performed in post processing.

In our framework, we assume that the PDF of the $L/L_{\rm Edd}$ time series corresponds to the ERDF of the AGN population (as discussed in Section 2.1.1). We therefore implemented the algorithm in such a way that it allows to choose between different ERDF shapes proposed in the literature, such as a broken power-law or log-normal distributions (see, e.g., Caplar et al. 2015; Weigel et al. 2017; Caplar et al. 2018). We note that our fundamental assumption that the PDF corresponds to the ERDF is physically acceptable only if the simulation is long enough for the AGN to span the whole ERDF. Although the current knowledge about AGN variability does not allow to constrain this timescale yet, we assume that it should be on the order of $\sim 10^6$ yr (see Section 4.1). On the other hand, in order to compare the predictions to observations, the simulations require a time resolution of years, and preferably even shorter timescales. As a consequence, we need to simulate light curves with $> 10^7$ steps. This makes the simulations extremely challenging from a computational point of view. In order to maximise the performance and the length of the simulated light curves we thus implemented our entire algorithm to run on an architecture of graphical processing units (GPUs; Section 3.1).

#### 2.1.1. *The Eddington ratio distribution function as the variability PDF*

As mentioned above, we propose to use the Eddington ratio distribution function (ERDF) as the PDF of the AGN (population) variability. By construction, the ERDF describes the distribution of $L/L_{\rm Edd}$ among the galaxy population, i.e. the fraction of galaxies in a given $L/L_{\rm Edd}$ range, for a given sample of galaxies at a given redshift range[2]. Assuming that the ERDF shape does not vary significantly during the considered time span, and that all the galaxies in the considered sample have the same variability properties, the ERDF can also be interpreted as the distribution of $L/L_{\rm Edd}$ that a galaxy can have during this time span (up to a normalisation factor): a galaxy moves across the ERDF, spending more time at $L/L_{\rm Edd}$ states corresponding to higher ERDF values[3].

Many studies attempted to infer the ERDF directly from observations (e.g. Kollmeier et al. 2006; Kauffmann & Heckman 2009; Schulze & Wisotzki 2010; Aird et al. 2012; Jones et al. 2016). However, directly measuring the ERDF is challenging mainly due to selection effects and difficulties in measuring black hole masses $M_{\rm BH}$ for statistically significant samples (e.g. Trakhtenbrot & Netzer 2012; Shen 2013; Peterson 2014; Mejía-Restrepo et al. 2016). Caplar et al. (2015) and Weigel et al. (2017) (W17 hereafter) proposed a different approach, where they assume a simple ERDF shape and apply forward modelling to deduce the parametrisation of the ERDF by deconvolving the AGN luminosity function. Specifically, W17 determined the the ERDF for local AGN, while Caplar et al. (2018) showed that the typical $L/L_{\rm Edd}$ of the (broken power-law) ERDF increases with redshift following $\lambda^*(z) \propto (1+z)^{2.5}$ (at least to $z \sim 2$).

Different parametrisations have been proposed in the literature to describe the observed ERDF. The most common are the broken power-law (e.g. Caplar et al. 2015; Bongiorno et al. 2016; W17; Caplar et al. 2018) and the log-normal distributions (Kollmeier et al. 2006; Conroy & White 2013; Kauffmann & Heckman 2009). Since the aim of our study is to reproduce the *observed* AGN variability, and therefore the behaviour of (mainly) radiatively efficient AGN, we consider as local ($z \sim 0$) PDF the (normalised) ERDF inferred by W17[4] (see below). In later stages of the present work, we will use the analysis of Caplar et al. (2018) to evolve the ERDF to higher redshift. We stress again that these assumptions can be made only if the length of the simulated light curve is comparable to the time needed for the AGN to span the whole ERDF, and if the ERDF is not expected to vary significantly on these timescales.

---

[2] We note that in many empirical studies the ERDF is determined for the AGN population instead of the total galaxy population.

[3] In a monotonically declining ERDF, this translates to spending more time at lower $L/L_{\rm Edd}$ values than at higher ones.

Following W17, the broken power-law ERDF for local AGN is parameterised as:

$$\xi(\lambda) = \frac{dN}{d\log\lambda} = \xi^* \times \left[\left(\frac{\lambda}{\lambda^*}\right)^{\delta_1} + \left(\frac{\lambda}{\lambda^*}\right)^{\delta_2}\right]^{-1} \quad (1)$$

with parameter values $\log\lambda^* = -1.84$, $\delta_1 = 0.47$, $\delta_2 = 2.53$ and $\log\xi^* = -1.65$. On the other hand, the log-normal parameterisation is:

$$\xi(\lambda) = \frac{dN}{d\log\lambda} = \frac{\xi^*}{\tilde{\sigma}\sqrt{2\pi}} \times \exp\left(\frac{-(\log\lambda - log\lambda^*)^2}{2\tilde{\sigma}^2}\right) \quad (2)$$

where $\log\lambda^* = -3.25$, $\tilde{\sigma} = 0.64$, $\log\xi^* = -0.77$.

It is important to note that the two parameterisations are almost equivalent for the $L/L_{\rm Edd}$ range probed in W17, i.e. $\log(\lambda) \approx (-4) - (+1)$. However, the currently available data do not allow to constrain the ERDF shapes at lower ER values where the two parameterisations differ significantly, with the log normal distribution leading to fewer low $L/L_{\rm Edd}$ AGN (Fig. 1). A drop at low $L/L_{\rm Edd}$ is however expected since the ERDF cannot increase indefinitely at the lower end. In order to allow the future users of our implementation to use various ERDF shapes (as proposed in the literature), we have implemented in our simulations code the possibility to use both the broken power-law and the log normal PDF parametrisations (see Section 3.1). We note that in the simulations we will sample the PDF uniformly in linear space (as opposed to logarithmic space). Therefore, we will consider $dN/d\lambda$ instead of $dN/d\log\lambda$ (see 3.1 for more details).

---

[4] In the present work we are considering the values for radiatively efficient AGN.

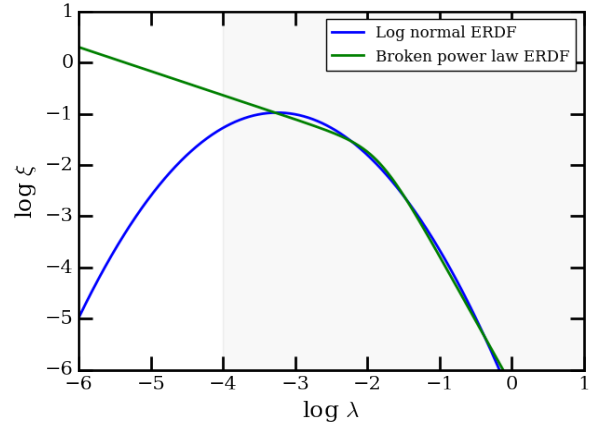

**Figure 1.** Eddington ratio distribution functions for local AGN, from W17, described either as a lognormal (blue) or broken power-law (green) distribution. The two descriptions are essentially indistinguishable for the $L/L_{\rm Edd}$ ($\lambda$) range probed in that work, i.e. $\log(\lambda) \approx -4 - 1$ (lightly shaded region). We note that while here we present $\xi(\lambda) = dN/d\log\lambda$ (i.e., space density per logarithmic unit in $L/L_{\rm Edd}$), as parameterised in Eqs. 1 and 2, our implementation uses a linear scaling, i.e. $dN/d\lambda$.

### 2.1.2. *Power Spectral Density*

The second statistical function needed from our simulations is the PSD that describes the time variability of $L/L_{\rm Edd}$. The total PSD can be interpreted as a superposition of different physical processes driving the variability at different timescales, and therefore at different frequencies, and its overall shape is not yet fully understood. Multiple studies have performed detailed investigations of the PSD on hours to decades timescales derived from light curves in the X-rays (in counts units, e.g. Markowitz et al. 2003; McHardy et al. 2004; González-Martín & Vaughan 2012; Paolillo et al. 2017) or optical/UV (in magnitude units, e.g. C17; Smith et al. 2018). Specifically, in the case of X-rays, the PSD is generally well described as a power law with slope approximately $\alpha = 2$, consistent with a random walk, at the highest probed frequencies (e.g. Lawrence & Papadakis 1993; Green et al. 1993), and one or two breaks leading to shallower low-frequency PSD (e.g. Uttley et al. 2002; McHardy et al. 2007). The break frequency appears to be correlated with BH mass and $L/L_{\rm Edd}$ (McHardy et al. 2004; McHardy et al. 2006), but the physical reason of this correlation is still unclear (e.g. McHardy et al. 2006; González-Martín & Vaughan 2012). On the other hand, studies based on different optical/UV photometric surveys (e.g. Zu et al. 2013; Caplar et al. 2017) or on *Kepler* light curves (Mushotzky et al. 2011; Kasliwal et al. 2015;

Smith et al. 2018) revealed steeper high-frequency PSD, strongly varying between different AGN.

The studies mentioned above all refer to PSDs in counts or magnitude units, while to the best of our knowledge no measurement or prediction for the PSD in $L/L_{\rm Edd}$ units is present in the literature. However, for the sake of simplicity, and inspired by the results above, we assume that also the PSD of $L/L_{\rm Edd}$ has a broken power-law shape. In this scenario, the bending may be associated to some specific physical process which suppresses the variability on progressively short frequencies, e.g. the response of the accretion disc to instabilities (e.g. Suberlak et al. 2017 and references therein).

Through the paper, we adopt the following parameterisation of a broken power-law:

$$PSD(f) = A \times \left[ \left(\frac{f}{f_{\rm br}}\right)^{\alpha_{\rm low}} + \left(\frac{f}{f_{\rm br}}\right)^{\alpha_{\rm high}} \right]^{-1} \quad (3)$$

where $f_{\rm br}$ is the break frequency, and $\alpha_{\rm low}$ and $\alpha_{\rm high}$ are the slopes at lower and higher frequencies, respectively (longer and shorter timescales, respectively). This parameterisation can however be modified to add additional breaks if needed. In addition, we do not consider any evolution of the variability behaviour, and therefore of the PSD, with redshift, as supported by some observations (e.g. C17; Paolillo et al. 2017). We stress that throughout the paper we will often refer to the periodogram, which is the statistical estimator of the PSD. The periodogram can be directly computed from a time series, and scatters around the underlying PSD (see Appendix A).

### 2.2. Converting Eddington ratio time series to observables: light curves and structure function

As discussed in the Introduction and at the beginning of this section, the primary output of our simulations are time series in units of $L/L_{\rm Edd}$. In order to compare the simulations to observations, or to make predictions for future surveys, the simulated $L/L_{\rm Edd}$ time series have therefore to be converted to observables. First, the simulated $L/L_{\rm Edd}$ time series can be converted to light curves in luminosity or magnitude units at the same wavelength range as the observations we want to compare to. The forward modelling approach allows to test different conversion recipes, e.g. by considering constant or luminosity dependent bolometric corrections (e.g. Marconi et al. 2004 and references therein), as well as introducing flux limits or other observational biases proper of every observation. The obtained light curves can then be treated as real observations and used to compute other quantities characterising the AGN variability behaviour. Specifically, in this work we concentrate on the (ensemble) structure function (SF), which quantifies the characteristic amount of variability for light curve (or in general time series) measurements separated by a given time interval $\tau$. A detailed definition and discussion of the SF is given in Appendix A (see in particular Eq. A6).

### 3. SIMULATIONS

As discussed in the previous sections, in this paper we present a new approach to investigate AGN variability on multiple timescales and among different objects which is based on the simulation of $L/L_{\rm Edd}$ time series. In this Section we provide details about the implementation of our simulations code, and we illustrate the extensive tests performed in order to characterise both the intrinsic and the numerical behaviour of the simulations.

#### 3.1. Code Implementation

We implemented the algorithm proposed by E13 to be executed on GPUs. Specifically, we wrote our implementation in C++ within the CUDA programming framework[5], with the libraries cuFFT[6], cuRAND[7] and Thrust[8]. This platform choice allows us to maximise the length (in steps) of the simulations, and is therefore more suited to our science goal than other available implementations based on, e.g., MATLAB (E13) or Python (Connolly 2015). This is particularly due to the fact that the algorithm requires multiple (inverse) discrete Fourier transform (DFT) and sorting operations[9], which are time consuming especially on arrays of the lengths considered in our work (up to $\sim 10^8$ points), and are being optimised on this platform.

We note that the current implementation is optimised for broken power-law and log-normal PDFs, and for single- or broken power-law PSDs. Specifically, for the random draw step we apply slice sampling (Neal 2003) from a normalised ERDF that is parameterised as in Eqs. 1 and 2. This choice of PDF and PSD shape is motivated by the observed ERDF and PSD, as discussed in Section 2. However, additional functional forms can be added if needed. As mentioned in Section 2.1.1, we are sampling the PDF uniformly in linear space.

#### 3.2. Testing and characterisation of the simulations code

---

[5] Version 9.0, https://developer.nvidia.com/cuda-downloads
[6] https://docs.nvidia.com/cuda/cufft/index.html
[7] https://docs.nvidia.com/cuda/curand/index.html
[8] https://docs.nvidia.com/cuda/thrust/index.html
[9] As an example, 1 simulation with $N_{\rm steps}$ iterations requires $O(N)$ DFT, $O(N)$ IDFT and $O(N)$ sorting operations. See also Section 3.2.3

We tested and optimised our code for the TITAN X (Pascal) NVIDIA GPU. In the following we describe a series of tests that we performed to characterise both numerical effects and intrinsic behaviour of the simulations.

#### 3.2.1. *Distortions in the Timmer & König algorithm*

The first step of the E13 algorithm consists in running the TK95 algorithm to create a time series of the desired length with a periodogram consistent with the input PSD. The PDF of the output time series should asymptotically approach a normal distribution for increasing number of steps (e.g., E13). However, we found that this is not true for power-law PSDs with steep slopes, $\alpha \gtrsim 1.5$ (PSD $\propto \nu^{-\alpha}$). We illustrate these distortions for several choices of $\alpha$ in Fig. 2. We conducted a number of simple tests in an attempt to understand the origin of these distortions. We found that the deviation from a normal PDF does not depend on the number of steps, and is therefore not a consequence of the finite simulation length. We also found that the same behaviour is present in simulations performed with other open codes (e.g., astroML, Ivezić et al. 2014; Vanderplas et al. 2012; DELCgen, Connolly 2015) and is therefore not an artefact of our specific implementation. Since our simulations only rely on the PSD of the TK95 output, and not on its PDF, this discrepancy between claimed and observed PDF does not affect our results. However, the TK95 algorithm is extensively used in the literature, and this aspect should be further investigated. We therefore decided to report it here for future reference.

#### 3.2.2. *Iterations and convergence*

The algorithm that we use to produce the time series is iterative in nature (E13, see also Section 2.1), and so has to converge in order to be robust. The algorithm that we use to produce the time series is iterative in nature (E13, see also Section 2.1), and so has to converge in order to be robust. As described in E13, while the values which can be assumed by the points in the simulated time series are determined at the beginning of the algorithm and do not change during the iterative process, these values are reordered at every iterative step until their variability converges to the variability described by the input PSD. Following E13 and S96, we consider this convergence is reached when the order of the time series values does not (significantly) change anymore between two consecutive steps. The exact number of iterations needed for the algorithm to converge to the final time series depends on the total number of steps, the assumed PDF and PSD, as well as the seed used for the random draw process. E13 proposed a convergence test where, at each step, the periodogram of the resulting time series is fitted with the same functional form as the input PSD, and the convergence of the fit parameters with respect to the input ones is considered. However, due to the significantly larger number of steps used in our simulations – more than 3 orders of magnitude larger than what considered in previous applications of the algorithm – this convergence test would be too time consuming to implement in our case. We therefore decided to base our convergence criterion on the ratio between the time series obtained from two consecutive iterations. Specifically, at every iteration $i$ we consider the following distance measurement $\delta_i$ between time the series $x_i(t)$ and $x_{i-1}(t)$:

$$\delta_i = \mathrm{rms}\left[\log\left(\frac{x_i(t_n)}{x_{i-1}(t_n)}\right)\right], \; n = 1, ..., \mathrm{num.\ steps} \quad (4)$$

The use of logarithm is motivated by the large dynamical range of the values in time series that are simulated using our observationally-motivated ERDF as PSD. The threshold corresponding to convergence strongly depends on the input PSD and PDF, and should therefore be chosen by the user depending on their simulation setup and the goals of the performed analysis. We ran several tests with 200 up to 1000 iterations and found, for the considered setups, the simulated time-series remain essentially unchanged, except for some minor differences at the highest frequencies probed. We caution and stress that using our feamework for simulating the most extreme cases of AGN variability (e.g. > 3 dex difference in < 1 yr) may require a simulation setup with an increased number of iterations to reach robust results. In what follows, however, we do not put particular focus on such extreme events, and instead demonstrate how the framework can be used to describe the broad phenomena of AGN variability seen in large, wide-field time domain surveys.

#### 3.2.3. *Performance*

The maximum number of steps per simulation allowed in the current implementation is $2^{27}$ ($\sim 10^8$), which is set by memory limitations[10]. In the astrophysical context of the present study, this may correspond to, e.g., a $\sim 13$ Gyr long simulation with $\sim 100$ yr resolution, or a $\sim 10^6$ yr long simulation with $\sim$ day resolution.

The running time is currently limited by the random draw and sorting steps of the algorithm (steps *ii* and

---

[10] The number of steps is defined in powers of 2 to maximise the simulation speed, in particular with regard to the discrete Fourier and inverse discrete Fourier transforms. All the arrays involved in the simulations are allocated in shared memory for the CPU and GPU using `cudaMallocManaged`.

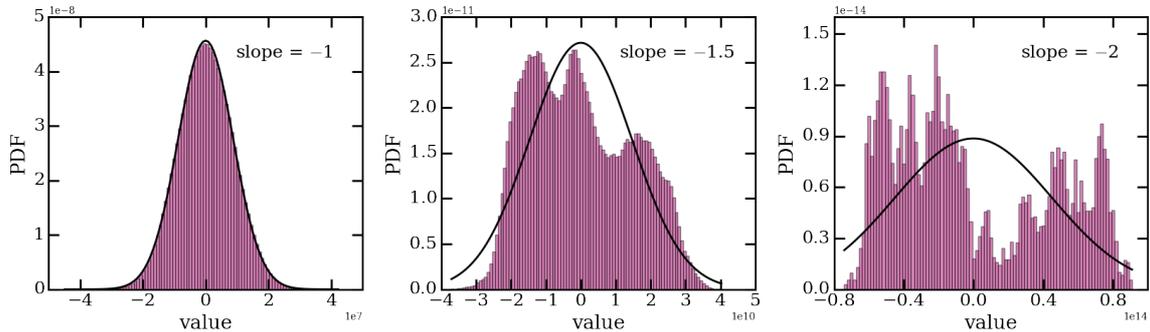

**Figure 2.** Example *output* PDFs of simulated time series based on the TK95 algorithm and on single power-law PSDs with various slopes (as indicated in each panel). In each panel, the black lines show normal distributions which match the output PDFs in mean values and in standard deviations. For steep PSDs, the resulting PDF deviates from a normal distribution. We note that, differently from the text, for this image we assume PSD $\propto \nu^{\alpha}$ instead of PSD $\propto \nu^{-\alpha}$

*iv* in Fig. 1 of E13, respectively). On the other hand, using cuFFT ensures that the (inverse) Fourier transform steps are performed in the most efficient way. Since the running time is mostly determined by the random draw step, it can vary depending on the assumed PDF shape. However, typical running times for 1 simulation with 200 iterations on TITAN X (Pascal) NVIDIA GPU machines do not exceed a few / several minutes.

### 3.2.4. *Dependencies on input parameters, combining and dividing light curves*

In this subsection we discuss how the final $L/L_{\rm Edd}$ time series, PDF, periodogram and SF shapes depend on the following input quantities:

- $t_{\rm bin}$: time separation between consecutive steps
- $N_{\rm steps}$: number of steps
- $T$: simulation length in the same units of time as $t_{\rm bin}$. This quantity combines $t_{\rm bin}$ and $N_{\rm steps}$, i.e. $T = t_{\rm bin} \times N_{\rm steps}$.
- $\mu$: the mean of the simulated $L/L_{\rm Edd}$ time series. For large $N_{\rm steps}$ this corresponds to the mean of the input PDF.
- $\sigma$: the standard deviation of the simulated $L/L_{\rm Edd}$ time series. For $N_{\rm steps}$ that is large enough this corresponds to the standard deviation of the input PDF.

In addition, we look at the effect of combining or dividing light curves obtained from different simulations. Indeed, given the challenge to run long simulations (see Section 3.2.3), a naive approach would have been to consider "stitching" together separate shorter simulations. On the other hand, understanding the behaviour of sub-samples is critical for the interpretation of observed light curves, which actually span only a short time window of the hole AGN life (or, in the case of these simulations, of the time needed for the AGN to span the whole ERDF). We note that in this sub-Section we refer to quantities computed in $L/L_{\rm Edd}$ units, while a discussion of the actual observables, i.e. magnitudes, is given in Section 3.2.10.

For these tests we considered multiple simulations with broken power-law PDFs inspired by the ERDF from W17 (see Eq. 1) and a broken power-law PSD as in Sartori et al. (2018a) (BPL3), but different choices of $t_{\rm bin}$ and $N_{\rm steps}$. A summary of the input parameters for the different simulations is given in Table 1. Specifically, we consider $L/L_{\rm Edd}$ time series simulated assuming different combinations of $t_{\rm bin}$, $N_{\rm steps}$ and $T$ (ER_sim_1 - ER_sim_6 in Table 1, we will refer to these simulations as "direct simulations"), as well as $L/L_{\rm Edd}$ time series obtained by dividing (ER_sim_2_cut) or stitching together (ER_sim_1_comp) direct simulations. We note that although some of the observed features depend on the specific adopted PSD and PDF, the main conclusions from these tests are valid also for different simulation setups.

Shape of the $L/L_{\rm Edd}$ time series

For the considered PSD and PDF set, the shape (global behaviour) of the $L/L_{\rm Edd}$ time series obtained from direct simulations (ER_sim_1 - ER_sim_6 in Table 1) is similar and does not depend on $t_{\rm bin}$, $N_{\rm steps}$ and $T$. Specifically, in this particular case all the simulations show one main prolonged period of elavated $L/L_{\rm Edd}$ (a "burst" or "switch on"), and a few additional, shorter "spikes". This is a direct consequence of the fact that each of the simulations spans the entirety of the same PDF, and of the (bent) power-law shape of the periodogram, which gives more power to the lowest frequencies (longer time separations).

The composite simulation (`ER_sim_1_comp`) shows more bursts, and therefore a higher level of variability on short timescales, compared to a direct simulation of same length (`ER_sim_2`, see Fig. 3). In fact, in the composite simulation every subsample has similar characteristic features (e.g. burst and spikes) as the direct simulation, although at different timescales. This means that stitching together shorter $L/L_{\rm Edd}$ time series will not conserve the variability behaviour proper of the input PSD (see also discussion below). As a consequence, simulating and stitching together shorter light curves, although computationally cheaper (see Section 3.2.3), is not a feasible way to obtain longer $L/L_{\rm Edd}$ time series with both $L/L_{\rm Edd}$ values and variability properties consistent with the assumed PDF and PSD.

Cutting a long simulation into subsamples (`ER_sim_2_cut`) also returns $L/L_{\rm Edd}$ time series with different shapes compared to direct ones with same final length (`ER_sim_3`, `ER_sim_4`). These are in general smoother as they are not covering the whole $L/L_{\rm Edd}$ range allowed by the PDF, but only a (consecutive) part of it.

In summary, combining or cutting simulations returns $L/L_{\rm Edd}$ time series with different shapes and variability features compared to direct simulations with same final length. This effect is visible also in the PDF and periodogram, as discussed below.

Probability density function

By construction, the PDF of every direct simulation (`ER_sim_1`-`ER_sim_6`), and of the composite simulation (`ER_sim_1_comp`), is statistically consistent with the input PDF (e.g. Fig. 4, left). In addition, as expected, the sampling of the PDF is more accurate for high values of $N_{\rm steps}$, especially at the increasingly high $L/L_{\rm Edd}$ end, which is sampled increasingly rarely. The lack of points with high $L/L_{\rm Edd}$ for low $N_{\rm steps}$ simulations has to be taken into account when looking at specific features in the $L/L_{\rm Edd}$ time series, which may ultimately be interpreted as extreme variability events (e.g., changing-look AGN).

On the other hand, the different subsamples (`ER_sim_2_cut`) also show PDFs which are (mostly) continuous in the $L/L_{\rm Edd}$ space, but these do not overlap in any way/range with the input PDF (Fig. 5, top). Specifically, most of the subsamples span a narrower $L/L_{\rm Edd}$ range. This is due to the fact that the $L/L_{\rm Edd}$ time series values are not randomly distributed (in time) but follow an order defined by the PSD. As a consequence, the PDF of light curves corresponding to different times during the AGN life can be significantly different.

Periodogram

By construction, all the periodograms computed for the direct simulations (`ER_sim_1` - `ER_sim_6`) have shapes consistent with the input PSD (e.g. Fig. 4, right). As described in Appendix A, the normalisation of the output periodogram differs from the input one, and depends on all the other aforementioned parameters, as well as on the covered frequency range. We verified that the normalisations of the periodograms computed for direct simulations are consistent with the expected value.

The periodograms computed for the subsamples (`ER_sim_2_cut`) show some distortions compared to the input PSD which are more important for subsamples with larger dynamical range in $L/L_{\rm Edd}$ (Fig. 5). A detailed discussions of these distortions, including both well known spectral distortions and effects related to the simulations, is given in Section 3.2.7 and Section 3.2.9. On the other hand, the periodogram of the composite simulation (`ER_sim_1_comp`) also shows some distortion which in this case may be explained by a suppression of variability at long timescales (every subsample has same $L/L_{\rm Edd}$ values, therefore not allowing the formation of a single peak as observed in the direct simulation, Fig. 3 top) and an increase of variability at short timescales (every subsample spans the whole PDF, Fig. 3 bottom).

Structure function

We construct SF following Eq. A6 and Sartori et al. (2018a). All the SF computed for direct $L/L_{\rm Edd}$ time series (`ER_sim_1`-`ER_sim_6`) show a broken power-law shape, which seems to reflect the PSD shape. Indeed, the two slopes coincide asymptotically (for $\tau \to 0$ and $\tau \to \infty$) to the prediction of the Wiener-Khinchin theorem[11] and the break is broadly consistent with $\tau_{\rm br} \sim 1/\nu_{\rm br}$.

In the case of composite and divided $L/L_{\rm Edd}$ time series (`ER_sim_1_comp`, `ER_sim_2_cut`), the SFs show a similar behaviour as described above for the periodogram, again confirming a relation between the two quantities. Specifically, distortions of the PSD at high frequencies are reflected to distortions of the SF at low $\tau$.

The normalisation of the SF differs between our various simulations. For a fixed PSD, it depends on the total length of the simulation ($T$), as well as on the standard deviation of the time series ($\sigma$). Specifically, for pure power-law PSD we empirically found that the normalisation has the following proportionality:

$$\mathrm{norm}_{\mathrm{SF}} \propto \frac{\sigma}{(t_{\mathrm{bin}} \times N_{\mathrm{steps}})^\beta}, \quad (5)$$

where $\beta = (\alpha - 1)/2$ is the same exponent as in the Wiener-Khinchin theorem, i.e. the exponent expected for the SF (in an ideal case). For broken power-law PSD, the difference between SFs obtained from different simulation setups (i.e. different $T$ and $\sigma$) is less well defined, as the ratio between two SFs changes as a function of $\tau$ (i.e., the SF have slightly different shapes). Although we were not able to find an analytical description of the normalisation in this case, we see similar dependencies: the overall SF is higher for increasing $\sigma$ or decreasing $T$, although the dependency on $T$ is more important for steeper PSD. This can be understood as follows. By construction, the time series obtained with our simulations span the entire PDF (i.e., the ERDF). If the simulation length is shorter, this means that the PDF has to be covered faster, which translates to an overall higher level of variability and thus a higher SF. Similarly, a larger $\sigma$ corresponds to a broader PDF, which, for fixed $T$ has to be again covered faster. This dependency of the SF normalisation on $T$ and $\sigma$ has to be considered when comparing simulations to observations, as discussed in detail in Section 4.1. Specifically, the length of the simulation has to be chosen such that the normalisation of the output SF (i.e. the SF of the simulated time series) matches the observed one. Moreover, we note that the normalisation does not depend on $t_{\mathrm{bin}}$ or $N_{\mathrm{steps}}$ separately, but on the combination of the two (i.e. simulations with different $t_{\mathrm{bin}}$ and $N_{\mathrm{steps}}$ but same $T$ have the same normalisation).

---

[11] The Wiener-Khinchin theorem states that a power-law PSD with slope $-\alpha$, $1 < \alpha < 3$ should corresponds to a power-law SF with slope $\beta = (\alpha - 1)/2$ (under specific assumptions; see also Emmanoulopoulos et al. 2010).

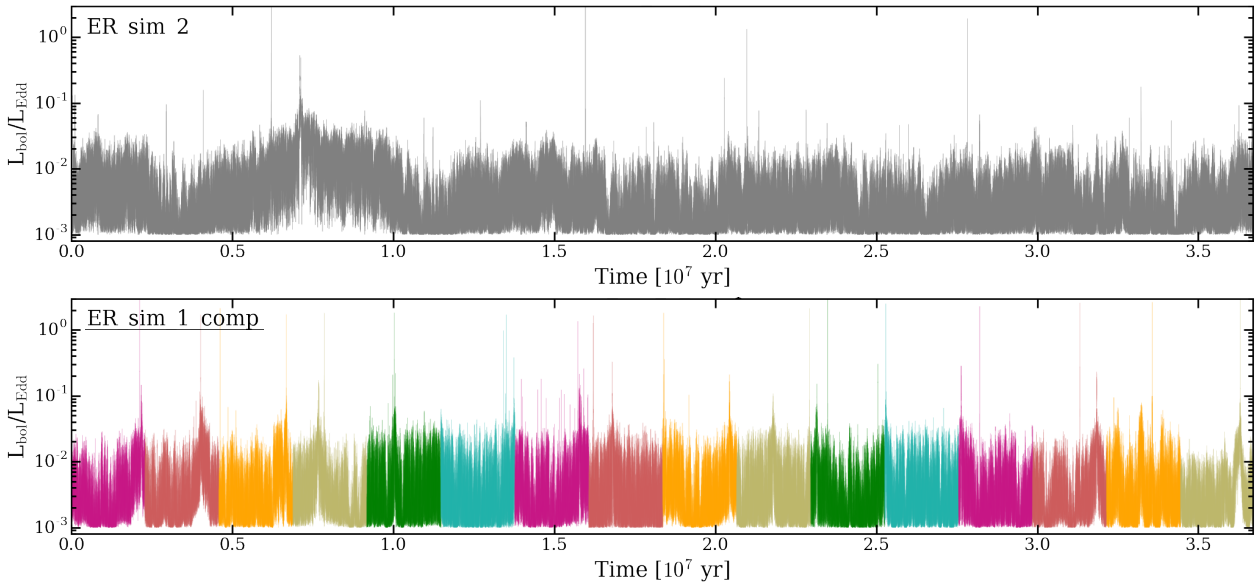

**Figure 3.** Testing the prospects of simulating increasingly long time-series. *Top:* $L/L_{\rm Edd}$ time series from a single simulation spanning the entire PDF (i.e., the entire ERDF; `ER_sim_2`). *Bottom:* composite $L/L_{\rm Edd}$ time series obtained by "stitching" together multiple shorter simulations (different colours), each of them also spanning the entire PDF (`ER_sim_2_comp`). Evidently, although both example data-sets span the same physical length (in yr) and $L/L_{\rm Edd}$ range, the direct (top) and composite (bottom) simulations present different features. Specifically, the composite simulations show multiple peaks (periods of elevated $L/L_{\rm Edd}$), approximately one per short simulation, while the single simulation only has one such peak (around $0.75 \times 10^7$ yr). This can be naturally understood given that each of the short simulations is forced to cover the entire ERDF. As a consequence, stitching together light curves is not a feasible solution to obtain a realistically long $L/L_{\rm Edd}$ time series (with a given computational setup), since it will not preserve the $L/L_{\rm Edd}$ time series shape

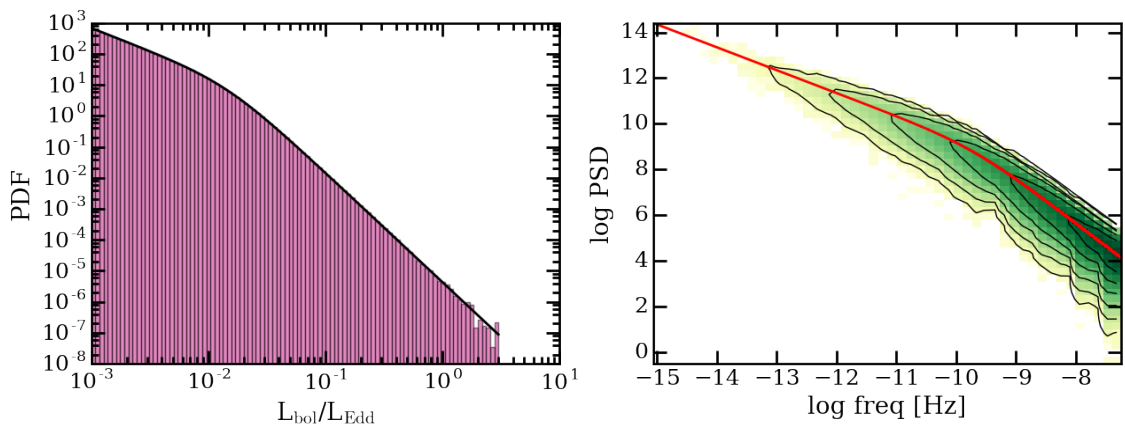

**Figure 4.** Comparison of input and output for a realistic choice of PDF (i.e., ERDF) & PSD (i.e., broken power-laws). *Left:* the PDF of the `ER_sim_2` simulation (purple) compared with the input PDF (black line). By construction, the PDF (i.e., ERDF) of the simulated time-series is statistically consistent with the input one. As expected, the sampling of the PDF becomes more accurate with increasing $N_{\rm steps}$, as is particularly evident at the high $L/L_{\rm Edd}$ end, where the sampling probability decreases. *Right:* the periodogram of the same simulation. The red line illustrates the input PSD, normalised according to $\mu$, $\sigma$, $t_{\rm bin}$ and $N_{\rm steps}$ of the simulation (see Eqs. A4 and A5). Here, too, the output periodogram is statistically consistent with the input PSD.

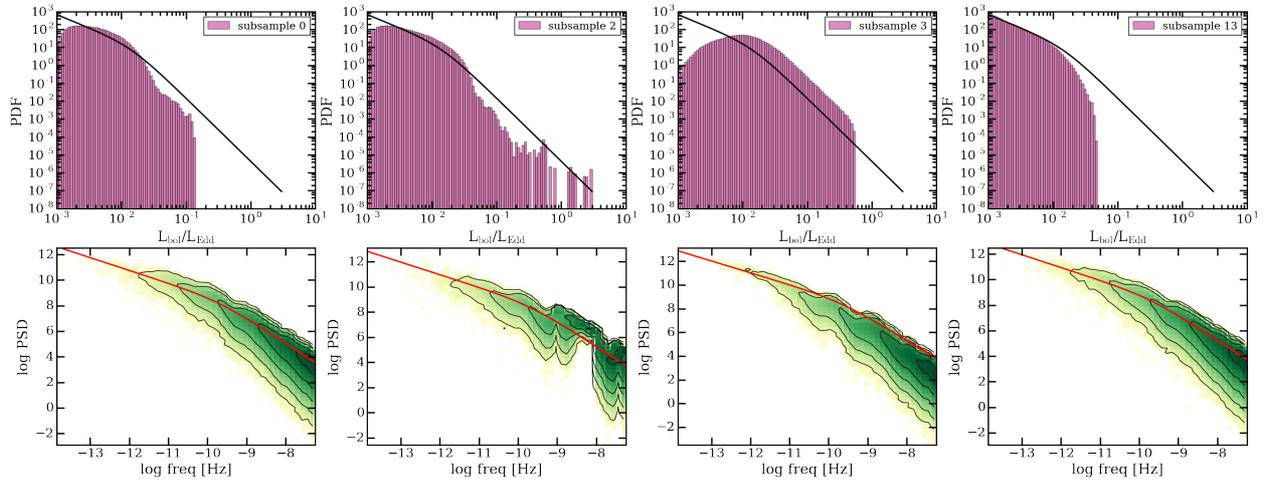

**Figure 5.** Same as Fig. 4 but for four sub-samples of `ER_sim_2` (from `ER_sim_2_cut`; that is, shorter simulations). The PDFs of the sub-samples do not match well the input PDF, as the simulated time-series are not sufficiently long to cover the entire PDF. Moreover, the periodogram show distortions, which are more important for subsamples covering a broader $L/L_{\rm Edd}$ range. See text for discussion.

3.2.5. *Mean variability behaviour from multiple realisations*

Since we are working with stochastic processes, every simulated ER curve will have a different periodogram and SF, i.e. they fluctuate around the input PSD and SF. By construction, at every frequency the periodogram scatters around the intrinsic PSD as defined by the Timmer & König algorithm (see Section 3 in Timmer & Koenig 1995 or Appendix A2 in E13 for more details). On the other hand, the SF of different realisations have similar shapes but different normalisations. Both effects are illustrated in Fig. 6.

In order to compare the simulated SF to observations as proposed in Sartori et al. (2018a), we need to define a representative SF and properly take into account the scatter among the different realisations. To get a representative SF for a given input set, $\mathrm{SF}(\tau)_{\mathrm{rep}}$, we therefore repeat the simulation $N_{\mathrm{sim}}$ times ($N_{\mathrm{sim}} > 50$; see discussion below), and compute the mean of the obtained $\mathrm{SF}(\tau)^2$ (see also C17):

$$\mathrm{SF}_{\mathrm{rep}}(\tau)^2 = \langle \mathrm{SF}(\tau)^2 \rangle = \frac{1}{N_{\mathrm{sim}}} \sum_{i=1}^{N_{\mathrm{sim}}} \mathrm{SF}_i(\tau)^2 \quad (6)$$

The standard error on $\mathrm{SF}(\tau)_{\mathrm{rep}}$ at every $\tau$ can then be estimated using the scatter in the set of calculated $\mathrm{SF}_i(\tau)^2$. Uttley et al. (2002) used a similar procedure to determine the intrinsic PSD from multiple light curves, and claimed that $> 50$ realisations are needed in order to reliably estimate the standard errors on the simulated periodograms. The same seems to also hold for the SF, but more testing should be done for every specific input set.

We note that in general the time span probed by observations are much shorter compared to the simulated ones. In order to probe that the observed SF can be directly compared to the representative one (i.e., there is no distortion or different normalisation), we first simulated 200 time series with $\sim 10^6$ points each and computed the representative SF (Fig. 6, right). From every time series we then randomly selected a (consecutive) subsample of 1000 points, obtained the corresponding SF, and then computed the representative SF (Fig. 7). Although the shorter SFs show a larger dispersion, the representative SF for the subsamples overlaps with the representative SF of the long simulations. This means that the representative SF can be directly compared to SF computed taking into account only subsamples (see also 3.2.6). We confirmed that holds by considering various input PSD shapes, and by computing the SF in magnitude units instead of $L/L_{\mathrm{Edd}}$ units.

3.2.6. *Ensemble analysis*

Since the SF is a statistical measurement, it is possible to compute it for a single source only in the case of a well sampled light curve (e.g. Suberlak et al. 2017 and references therein). However, for sparsely sampled light curves, the SF is usually computed in an ensemble way (e.g. Sesar et al. 2006; MacLeod et al. 2012; C17). In this case, it is assumed that all the sources in the considered sample have the same variability behaviour, and that the SF at a given time lag $\tau$ is computed with Eq. A6, by considering simultaneously all the pairs of data points separated by a corresponding time lag, in all the sources under question.

In order to test if SFs computed in an ensemble way (i.e. considering subsamples of different light curves) are consistent with SFs computed directly from individual simulations, we performed a test similar to what is described at the end of Section 3.2.5. First, we ran 1000 simulations with $\sim 10^6$ points each and selected a (consecutive) subsample of 1000 data points from each simulation. For every simulation, we then masked the subsamples to mimic sparse measurements and computed the ensemble SF. As an example, Fig. 8 shows the ensemble SF obtained by considering only 0.5%, 1% and 10% of every subsample, respectively (randomly selected points). As expected, the ensemble SF always scatters around the representative SF, but the scatter significantly decreases with increasing number of considered points. This means that the mean shape and the normalisation of the SF are conserved with ensemble analysis, but the accuracy strongly depends on the number of measurements.

|  | Name | $t_{\rm bin}$ [ks] | $N_{\rm steps}$ | $T$ [$10^6$yr] | Comment |
|---|---|---|---|---|---|
| **Initial simulation** | ER_sim_1 | 8'640 | $2^{23} \sim 8 \times 10^6$ | $\sim 2.30$ | |
| **Longer simulation but with same $t_{\rm bin}$** | ER_sim_2 | 8'640 | $2^{27} \sim 10^8$ | $\sim 36.75$ | $16\times$ longer in yr |
| **Append simulations** | ER_sim_1_comp | 8'640 | $2^{27} \sim 10^8$ | $\sim 36.75$ | Append 16 realisations of ER_sim_1 |
| **Cut long simulation** | ER_sim_2_cut | 8'640 | $2^{23} \sim 8 \times 10^6$ | $\sim 2.30$ | Cut ER_sim_2 in $16\times$ pieces with same length and binning as ER_sim_1 |
| **Change $t_{\rm bin}$ and $N_{\rm steps}$, same $T$** | ER_sim_3 | 4'320 | $2^{24} \sim 2 \times 10^7$ | $\sim 2.30$ | $2\times N_{\rm steps}$, $0.5\times t_{\rm bin}$ |
|  | ER_sim_4 | 2'160 | $2^{25} \sim 3 \times 10^7$ | $\sim 2.30$ | $4\times N_{\rm steps}$, $0.25\times t_{\rm bin}$ |
| **Change $t_{\rm bin}$ and $T$, same $N_{\rm steps}$** | ER_sim_5 | 2'160 | $2^{23} \sim 8 \times 10^6$ | $\sim 0.60$ | $0.25\times t_{\rm bin}$ and $T$ |
|  | ER_sim_6 | 34'560 | $2^{23} \sim 8 \times 10^6$ | $\sim 9.20$ | $4\times t_{\rm bin}$ and $T$ |

**Table 1.** Summary of simulations for the tests in Section 3.2.4. As a reference, 8'640 ks corresponds to 100 days.

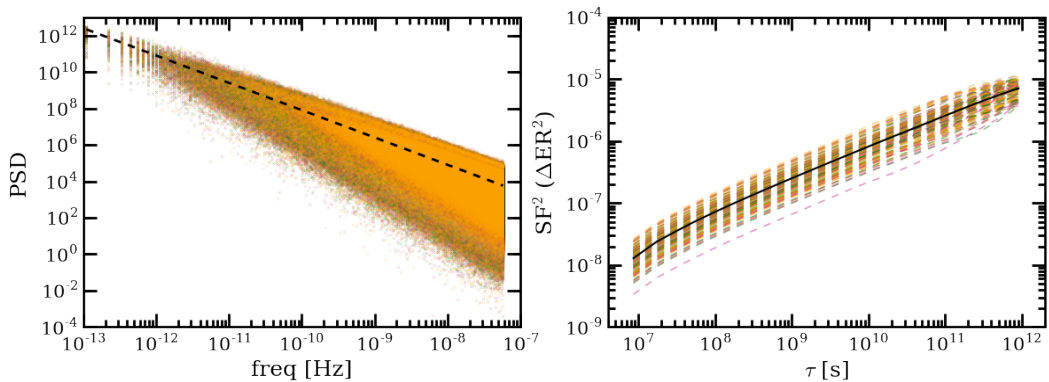

**Figure 6.** Scatter and robustness in repeated realisations of a time-series simulation. *Left:* periodograms for 200 realisations of the same simulation input with simple power-law PSD (slope $\alpha = 1.5$; coloured crosses). The black line traces the input PSD (renormalised, see Appendix A for discussion about PSD nomalisation). As expected, the periodograms measured from the simulated time-series scatter around the intrinsic (input) PSD. *Right:* SF$^2$ for the same simulations (dashed colour lines). The SF$^2$ measurements for different simulations have a similar shape, but differernt normalisations. The representative (mean) SF$^2$, defined following Eq. 6, is shown in black. We note that, as discussed in Section 3.2.8, some distortions can be observed at the longest probed timescales (i.e., lowest frequencies).

We stress that what reported in the test above is only indicative of a trend. In fact, the scatter in the ensemble analysis strongly depends on the number of considered sources, and on the length and sampling of the light curves measured for each source. Therefore, in order to determine how well an ensemble SF computed for a given survey represents the intrinsic one (i.e. if we can compare the ensemble SF with our simulations), the test above should be repeated by assuming the same source numbers observational cadence as in the considered survey.

3.2.7. *Spectral distortions: red noise leak and aliasing*

The periodograms computed for finite, discrete light curves are subject to spectral distortions known as red noise leak and aliasing (see below for a definition, and Uttley et al. 2002 for a detailed discussion). In fact, measuring a light curve $l(t)$ with a time sampling given by the window function $w(t)$, where $w(t) = 1$ during the observations and 0 elsewhere, corresponds to convolving the Fourier transform of the underlying light curve $F(f)$ with the Fourier transform of the window function $W(f)$:

$$l_{\rm obs}(t) = l(t) \times w(t) \quad \Rightarrow \quad F_{\rm obs}(f) = F(f) * W(f) \quad (7)$$

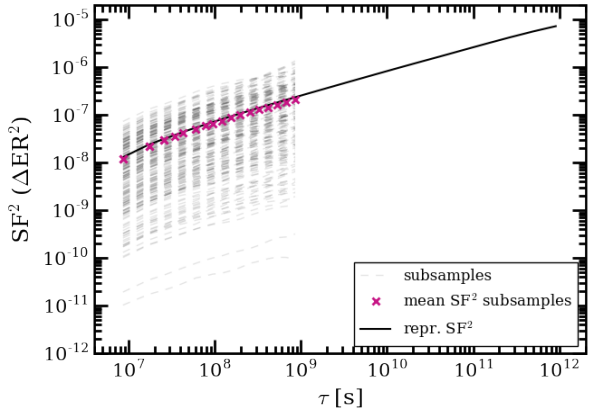

**Figure 7.** Robustness of the representative structure function. The SF$^2$ for the random sub-samples of the simulations shown in Fig. 6 (gray dashed lines, see text for more details). The representative SF$^2$ of the sub-samples (purple crosses) overlaps with the representative SF$^2$ of the long simulations (black line).

Since the periodogram is computed from $F_{\rm obs}(f)$ (see Appendix A), a distortion due to a window function is reflected as a distortion in the periodogram. Specifically, red noise leak (e.g. Deeter & Boynton 1982; Deeter 1984) appears when the length of the observed light curve is significantly shorter than the length of the intrinsic light curve, so that significant power is present at frequencies shorter than the ones probed with the observations. In this case, trends in the light curve due to frequencies below the observed limit (timescales longer than the light curve total length) are not distinguishable from trends at higher frequencies, and power is transported from low to high frequencies. This effect is more pronounced for steep PSDs (see Uttley et al. 2002 and references therein). On the other hand, aliasing (e.g. van der Klis 1997) appears when the time sampling does not allow to probe variability at high frequencies $f$, and in this case there is a fold-back of power from $f_{\rm Nyq} + \Delta f$ to $f_{\rm Nyq} - \Delta f$, where $f_{\rm Nyq}$ is the Nyquist frequency (see Appendix A). This effect is more pronounced for shallow PSD.

Both red noise leak and aliasing cause a flattening of the measured periodogram, at a level that dependents on the shape of the underlying PSD. Studies comparing simulated periodograms to observed ones should therefore include such distortions in the simulations. The standard way to do it is to simulate longer light curves (e.g 100 times longer, for red noise leak) with higher sampling (e.g. 10% of the observed sampling, for aliasing), and then randomly select a subsample with length and sampling as in the observations (e.g. Uttley et al. 2002).

### 3.2.8. *Distortions of the Structure Function*

While the spectral distortions discussed in Section 3.2.7 can significantly modify the computed periodogram, the SF is commonly thought to be less affected by finite, irregularly sampled light curves. However, Emmanoulopoulos et al. (2010) showed that spurious breaks can appear in the SF of single light curves even in the case of a power-law PSD (i.e. no intrinsic characteristic time-scales), with unpredictable behaviour at timescales longer than the break. The position of the break increases with light curve length and PSD steepness (see Fig. 5 in Emmanoulopoulos et al. 2010), so that the effect is less important in long light curves with steep PSDs. In addition, he showed that windowing of the light curve may also affect the precise SF shape (although the general behaviour is conserved). As can be seen e.g. in Fig. 6, some distortions at the longest probed timescales are present also in our simulations, although they are not as extreme as reported in Emmanoulopoulos et al. (2010). However, these distortions smooth out when considering the representative or ensemble SF, so that these distortions do not significantly affects statistical studies (see also Kozłowski 2017a; Guo et al. 2017), as it is the case in the applications presented in Section 4.

Although the distortions described in Emmanoulopoulos et al. (2010) do not affect ensemble studies, other distortions can be seen in our simulations and should be taken into account. According to the Wiener-Khinchin theorem, a power-law PSD with slope $-\alpha$, $1 < \alpha < 3$, should corresponds to a power-law SF with slope $\beta = (\alpha - 1)/2$ (under specific assumptions, see also Emmanoulopoulos et al. 2010). However, as can be seen in Fig. 9, for $\alpha \lesssim 1.5$ ($\gtrsim 2.5$) the obtained SF is steeper (shallower) than expected from the theorem. For a given time resolution, this effect is more pronounced for short timeseries, as the difference between the predicted and observed SF slope is higher at the lowest probed tau $\tau$, independently on the total number of steps (see Fig. 9 for TK95 simulations). The main reason of this discrepancy is that the observed and simulated light curves are not infinite and continuous in time, as required by the theorem. In addition, for steep PSDs we often see a distortion at the highest probed $\tau$ even in the representative SF. The main reason of these distortions is that the number of pairs for a given $\tau$ decreases with increasing $\tau$, and therefore the SF, which is a statistical measure, cannot be reliably computed for large $\tau$.

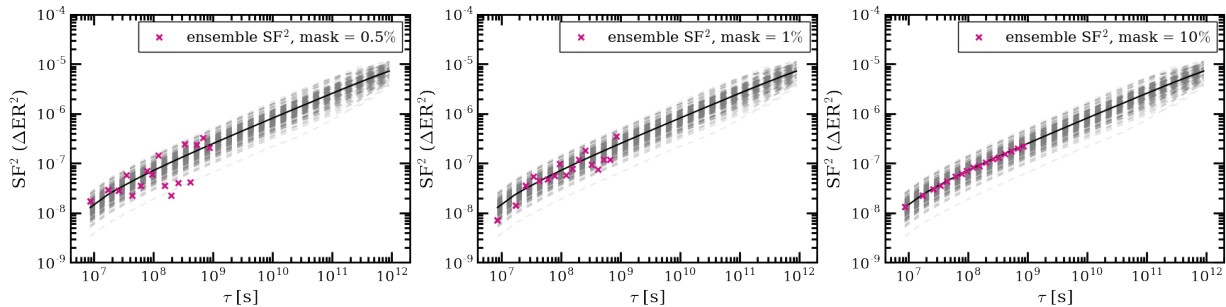

**Figure 8.** Ensemble structure functions for sub-samples of the simulated time-series – similar to Fig. 6 (right) and Fig 7, but for ensemble analysis. Here the purple crosses show the ensemble SF obtained by considering only 0.5% , 1%, and 10% (*right*) of the simulated data points in every sub-sample, (*left*, *centre*, and *right*, respectively; randomly selected points). The ensemble SF always scatters around the representative SF, but the scatter decreases significantly with increasing number of considered points.

The discussed issues strongly illustrate why one should be cautious when considering the highest $\tau$ SF measurements of any simulation, as these are essentially unreliable. The actual threshold above which the SF becomes unreliable depends on light curve sampling and length, as well as on the PSD shape. As a rule of thumb, in our analysis we are considering SFs only up to 1/10 of the total light curve length (e.g. C17). We stress, however, that for the PSD slopes expected from observations (see Section 4), and for ensemble studies, we do not expect our results to be biased by these distortions.

### 3.2.9. *PSD and SF collapse*

For some choices of PSD and PDF sets, the algorithm returns time series whose periodograms "collapse" towards utterly unrealistic PSD shapes. As an example, a PDF as described in W17 with $\lambda_{\min} = 10^{-5}$, $\lambda_{\max} = 10$ combined with a (broken) power-law PSD with low-frequency slope $\alpha \gtrsim 1.5$ leads to a collapse of the final periodogram to a line with negligible scatter and a slope close to $\alpha = 1$ (pink noise). This effect is observed for both lognormal and broken power-law PDFs, and the final shape mostly depends on the initial random seed (and not on other parameters such e.g. the input slope). An example of this collapse for 3 simulations with power-law PDF and 3 different slopes, $\alpha = 1.6, 1.8$ and 2.0, is given in Fig. 10. We note that a distortion of the periodogram at high (low) frequencies is usually reflected in a distortion of the SF at short (long) $\tau$.

As we elaborate below, our tests only showed PSD and SF collapses for simulations setups which considered overall steep PSD ($\alpha \gtrsim 1.5$, see Fig. 10), and/or large dynamical ranges for the values in the considered time series (e.g., $\gtrsim 6$ dex, see Fig. 11). The observed collapse in the periodogram, as well as the distortions in the SF, disappear in the following circumstances (see example in Fig. 11):

1. The overall slope is shallower. This will very likely always be the case in most of our analysis, since a steep low-frequency end would mean a steep long-timescale SF, and this seem not to be allowed by the observations;

2. The dynamical range $[\lambda_{\min}, \lambda_{\max}]$ for the PDF is narrower (e.g. 2 dex instead of 6 dex);

3. We only consider a subsample of the LC, e.g. 10%. We note that in most cases this also corresponds to a smaller dynamical range (as the time series does not have enough time to traverse the entire PDF);

4. We consider the SF in magnitude units instead of $L/L_{\rm Edd}$. We note that, since $\Delta {\rm mag} \propto \Delta \log L \propto \Delta \log L/L_{\rm Edd}$ (see below), working in magnitude instead of $L/L_{\rm Edd}$ results in a reduced dynamical range (i.e., a PDF covering 4 dex in $L/L_{\rm Edd}$ would be mapped to 10 magnitudes).

These conditions suggest that the observed distortions are due to a numerical issue (that is, the LC intrinsically have the right PSD and SF, but we are not able to compute it) instead of being intrinsic of the LC. Specifically, since the SF is computed by looking at the differences between two different points, it is numerically harder to compute it carefully when the dynamical range of the time series is larger (i.e. a roundoff error). In addition, it is possible that some choices of PSD+PDF sets are not allowed for mathematical reasons (as an extreme, a very sharp PDF would not allow almost any variability). Although we are not investigating this issue further, we acknowledge this effect and stress that this issue does not affect our results since the region of the parameter space where this sort of collapse occurs is not considered in our analysis. Indeed, as discussed in Section 2.1.2 and Section 4.1.2, the overall PSD is expected to break and flatten beyond a given frequency, such that the overall (mean) slope is shallow enough not to cause the collapse (see also Point 1 in the list above). In addition, the comparison to observed data is usually performed in magnitude (i.e., logarithmic) space instead of in linear $L/L_{\rm Edd}$ or luminosity space, which naturally reduces the dynamical range for the values in the considered time series, in a significant way.

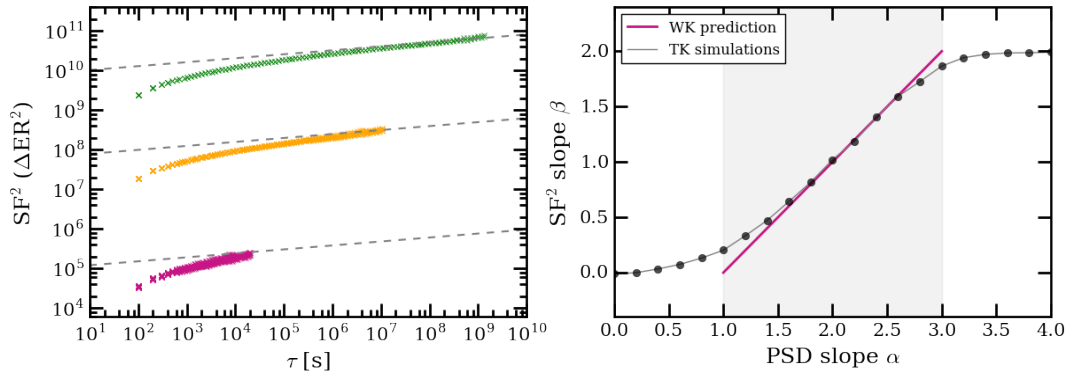

**Figure 9.** *Left:* Structure functions computed for TK95 simulation with PSD with a slope $\alpha = 1.1$, a fixed time step but different lengths. The grey dashed lines have slopes as expected from the Wiener-Khinchin theorem for this PSD (arbitrarily normalised). All the simulated SFs show a downturn at the shortest probed timescales, which is independent of the simulation length. *Right:* resulting SF slopes as a function of input PSD slope (black points). Each point was obtained by fitting a line to 100 TK95 simulations (for every PSD slope). Here SF $\propto \tau^\beta$ and PSD $\propto \nu^{-\alpha}$. The results deviate from expectation from Wiener-Khinchin theorem (purple) as the PSD slopes deviates from $\alpha = 2$, with higher (lower) mean SF slopes for decreasing (increasing) PSD slopes.

### 3.2.10. *Structure function in Eddington ratio vs. structure function in magnitude*

As we mentioned in Section 2.2, the primary output of our simulations are $L/L_{\rm Edd}$ time series which, in order to be compared to observations, have to be converted to light curves. Specifically, most of the ensemble studies considered in this work compute SFs in *magnitude* units, $SF_{\rm mag}$ (e.g. de Vries et al. 2003; Sesar et al. 2006; MacLeod et al. 2010; MacLeod et al. 2012; C17). A description of how to convert from $L/L_{\rm Edd}$ to luminosity curves is given in Section 2.2. However, in the simplest case where the luminosity $L(t)$ in the considered band is proportional to the Bolometric luminosity $L_{\rm bol}(t)$[12],

---
[12] We assume that the BH mass is not increasing significantly during the time probed by the simulations.

the magnitude difference needed to compute the $SF_{\rm mag}$ (eq. A6) can be directly derived from the $L/L_{\rm Edd}$ time series following: $m(t_2) - m(t_1) = -2.5 \log \left[\frac{L(t_2)}{L(t_1)}\right] = -2.5 \log \left[\frac{L_{\rm bol}(t_2)}{L_{\rm bol}(t_1)}\right]$

$= -2.5 \log \left[\frac{\lambda_{\rm Edd}(t_2)}{\lambda_{\rm Edd}(t_1)}\right]$ where $\lambda_{\rm Edd}(t)$ is the Eddington ratio at time $t$ ($L/L_{\rm Edd}$ throughout the text).

By inspecting multiple simulations we found that, if no distortions are present in $SF_{L/L_{\rm Edd}}$ (see Section 3.2.9), then for the same simulation the $SF_{\rm mag}$ computed following equations A6 and 3.2.10 would have a similar shape, but different normalisation, compared to $SF_{L/L_{\rm Edd}}$. On the other hand, for the cases where $SF_{L/L_{\rm Edd}}$ shows distortions due to, e.g., a large dynamical range of the PDF (ERDF), these distortions disappear for $SF_{\rm mag}$. The true underlying reason for this is not yet clear.

## 4. APPLICATIONS

In the previous sections we presented our framework to simulate AGN light curves starting from the distribution of $L/L_{\rm Edd}$ among the galaxy population. In the following we illustrate some early science applications for the framework, and describe how this framework (and, hopefully, similar studies pursued by the community) will become increasingly important to best exploit new observational data and surveys.

We note that the observational data available to date, and in particular the lack of observational constraints on super-human timescales, do not allow to fully apply our framework to constrain the underlying PDF+PSD driving AGN variability (or to probe its existence). How-ever, new IFU observations of galaxies with extended AGN photoionised gas (e.g. the *Voorwerpjes* galaxies, Lintott et al. 2009; Keel et al. 2012a; Keel et al. 2012b; Schawinski et al. 2010; Gagne et al. 2014; Sartori et al. 2016; Keel et al. 2017; Sartori et al. 2018b) will allow us to reconstruct hystorical AGN ligth curves (e.g. Treister et al. 2018) and gain new information about AGN variability on $> 10^4$ yr timescales. This will be crucial to constrain our models including data at timescales which are still poorly understood. In addition, new time domain surveys, such as the upcoming LSST, will provide light curves for millions of sources (LSST Science Collaboration et al. 2009) which will allow us to better investigate the variable behaviour of single AGN, as well as compare it to the ensemble behaviour of the general AGN population (e.g. MacLeod et al. 2010; MacLeod et al. 2012; C17).

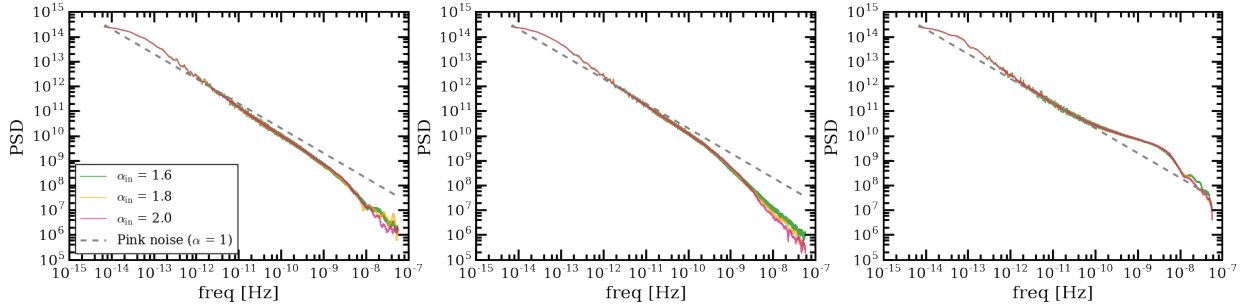

**Figure 10.** Example of the periodogram "collapse". We show three sets of simulations with broken power-law PDFs, and power-law PSDs with slopes of $\alpha = 1.6$, 1.8, and 2.0 (green, orange, and purple, respectively). The three panels correspond to three initial random number generator seeds. The periodograms collapse to a line with negligible scatter and slope close to $\alpha = -1$ (pink noise), whose specific shape mostly depends on the initial random seed.

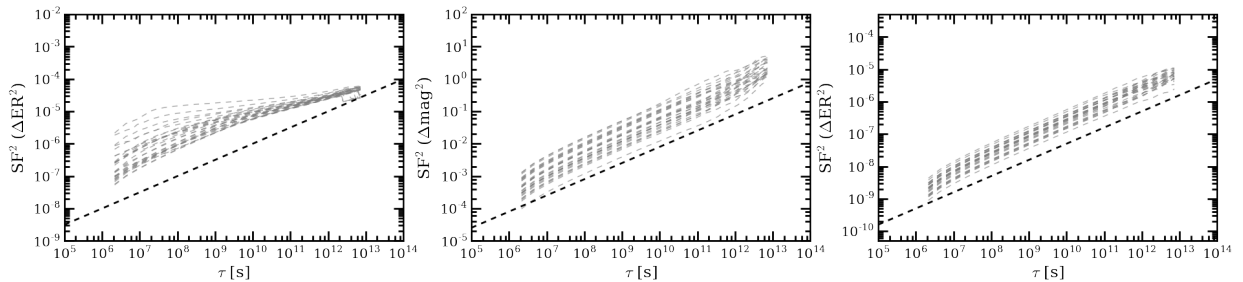

**Figure 11.** Distortions of the structure function due to dynamic range of the PDF. *Left:* SF computed for 30 $L/L_{\rm Edd}$ time series simulated with the same input parameters (grey dashed lines), i.e. a power-law PSD with slope $\alpha = 1.5$ and a broken power-law PDF spanning 6 dex in $L/L_{\rm Edd}$. The SFs are distorted compared to the predictions of the Wiener-Khinchin theorem (black dashed line, arbitrarily normalised). The distortions are clearly reduced if the SF is computed in magnitude instead of $L/L_{\rm Edd}$ (*centre*), or if the PDF covers a narrower dynamical range (*right*, here 1 dex).

### 4.1. *Constraining specific models with existing AGN variability data*

The goal of this section is to provide an example of how existing AGN variability data, in this case the optical SF measurements from the PTF/iPTF survey (Bellm 2014) presented in C17, can be used to constrain the input model parameter space for the kind of simulations enabled by our framework. Specifically, we want to find a simulation setup such that the derived 4000 Å (*R*-band) SF is consistent with the observed one, presented in C17.

As described in Section 3.2, the shape and normalisation of the simulated SF depend on the input PSD and PDF, while the simulation's length in physical units ($T$, e.g. yr) also influences the overall normalisation (in particular for steep PSDs). Since different normalisations correspond to different variability behaviours, it is crucial to choose physically motivated assumptions for these input parameters in order to obtain realistic SFs to use for our analysis. In addition, observational biases such as flux limits and sample selection can influence the shape and normalisation of the observed SF, and therefore have to be included as additional steps in the simulation process[13]. In the following we outline how to use available observations to define a reference simulation setup consistent with the PTF/iPTF SF, and which will be used for other example applications in the next sections.

### 4.1.1. The PTF/iPTF structure function

C17 presented the ensemble 4000 Å SFs for a homogeneous sample of $\sim 28{,}000$ quasars selected from SDSS-DR7 (Shen et al. 2011) divided into 64 $M_{BH} - z - L_{bol}$ bins (median $z \sim 1.3$, median $\log(M_{BH}) = 9.1$; see Fig. 4 in C17). In order to obtain a representative SF for the whole PTF/iPTF sample, we divided the SF values C17 in 6 time bins and took the mean of all $SF^2$ measurements in that bin (see Fig. 13, black points). In the following we will refer to this representative SF as the PTF/iPTF structure function, $SF_{PTF,C17}$.

### 4.1.2. The input PDF and PSD

Following Section 2.1.1, we defined the input PDF as a broken power-law with slopes as in W17 (see Section 2.1.1) and $\lambda^* = 0.11$ as expected for $z \sim 1.3$ and the redshift evolution in Caplar et al. (2018). For these simulations we fixed $\log(\lambda_{max}) = 0.5$ in order to allow for short periods of super-Eddington accretion (and thus, super-Eddington sources). On the other hand, we chose $\log(\lambda_{min}) = -3.75$ such that the active fraction, defined as the fraction of time the AGN spends above $\lambda = 0.01$, is $\sim 10\%$ (this is broadly consistent with observations, e.g. Aird et al. 2018). We stress that different definitions of the active fraction are present in the literature, and that its value is not fully constrained yet, as it may depend on multiple parameters such as black hole mass, luminosity and redshift.

While the PDF can be motivated by observational constraints (e.g. ERDF, AGN fraction), the shape of the PSD is well less defined (see discussion in 2.1.2). However, inspired by the PSD obtained e.g. in mag or flux units we assumed a broken power-law shape with slopes and break to be determined through forward modelling, i.e. by comparing the SFs obtained through our suite of simulations to the observed one, $SF_{PTF,C17}$.

### 4.1.3. Creating mock ensemble structure functions

For every input PDF and PSD set (see Section 4.1.4 for the assumed inputs, including total simulation length), we ran 300 simulations (i.e. 300 realisations of the same input set) and created a mock ensemble SF to be compared to $SF_{PTF,C17}$ as follows. First, for every simulation we looked for one subsample with similar length as the PTF survey (assumed here to simply be 2000 days in the observed frame), and similar observational biases and sample selection as the sample selection in C17, following the procedure below:

1. *SDSS selection*: select a random point in the $L/L_{Edd}$ time series and convert $L/L_{Edd}$ to the observed SDSS $i$-band magnitude assuming $z = 1.3$ and $\log(M_{BH}) = 9.1$. Specifically, we convert $L_{bol}$ to observed magnitude by assuming the Vanden Berk et al. (2001) quasar composite UV-optical SED, shifted to $z = 1.3$, sampled at 3000 Å (rest frame) and applying a bolometric correction of $L_{bol}/\lambda L_\lambda(3000\,\text{Å}) = 3.25$ (e.g. Trakhtenbrot & Netzer 2012). Whenever the flux exceeds the SDSS main quasar sample flux limit, $i < 19.1$, we proceed to step *(ii)*; otherwise, we select another random point and repeat the calculation. This step assures that the mock AGN would have been selected within the main SDSS quasar sample, and therefore part of the initial sample used in C17.

2. *PTF sample*: take a 2000 days long subsample starting 5 yrs after the SDSS point[14].

3. *PTF masking*: convert $L/L_{Edd}$ to the observed PTF $R$-band magnitude (following the same conversions as in step (i) above) and apply a mask such that 1) all the points with $R < 20$ are observed, 2) all the points with $R > 21$ are not observed, and 3) the points with $20 < R < 21$ have a probability to be observed which is a linear interpolation in mag between 20 and 21. This step aims at reproducing the observational limits in the PTF sample with a simple prescription.

4. *PTF selection*: if the median of the masked subsample is $R < 19.1$, accept the subsample as a legitimate mock PTF light curve, otherwise go back to step *(i)*. This magnitude cut reproduces the specific sample selection applied in C17. We note that at the considered redshift ($z = 1.3$) and BH mass ($\log(M_{BH}/M_\odot) = 9.1$), the $R = 19.1$ magnitude cut corresponds to an $L/L_{Edd}$ cut of roughly $\log(L/L_{Edd}) = -1.2$, which is consistent with what seen in the C17 sample. As an illustration, Fig. 12 shows the regions above the PTF limit for 100 of the 300 simulations, as well as the position of a possible PTF region.

---

[13] We stress that the simulations presented and discussed in the previous sections did not consider sample selection or flux limits.

[14] We chose 5 yr as representative of the gap between SDSS and PTF observations but we tested that varying this gap by a couple of years does not significantly affect our results.

We then computed the ensemble *observed* SF, in magnitudes, from the 300 mock PTF light curves (i.e., observed $R$-band magnitudes), and converted it to *intrinsic* 4000 Å SF using Eq. 4 in C17. For every choice of input parameter set (i.e. every PDF and PSD set), we repeated this procedure 500 times (i.e. we computed 500 ensemble SF, each time using different subsamples), and computed the representative $SF_{PTF}$ following Eq. 6. We note that this is essentially equivalent to computing the ensemble SF for $300 \times 500 = 150'000$ AGN.

#### 4.1.4. *The reference simulation*

In principle, what described in Section 4.1.3 should be repeated following an MCMC approach by stepping over a grid of input parameters (i.e., PSD break and slopes, and simulation length), in order to find the best possible model for the data. However, as pointed out above, the current data are not sufficient to fully constrain the PSD, especially due to the lack of measurements beyond human timescales. Therefore, for now we can only use a forward modelling approach to test if a given set of input parameters is consistent with the observed ensemble analysis. For the PDF described in 4.1.2, we found that a PSD with $\alpha_{\rm low} = 1$, $\alpha_{\rm high} = 2$, and $f_{\rm br} = 1 \times 10^{-9}$ (~30 yr), and a simulation length of 1.4 Myr is consistent with the observations, as shown in Fig. 13. The reference simulation setup is summarised in Table 2, and will be used in the next Sections.

We stress that timescales on the order of the PSD break (~30 yr in intrinsic frame) or longer are not directly probed by the subsamples used for the ensemble analysis, as the iPTF/PTF data span only 5.5 yr. However, they *are* probed by the total light curves from which the subsamples are selected. Therefore, they indirectly affect also the timescales probed in the ensemble analysis. Specifically, in this case we saw that the break position influences the SF normalisation at the timescales that are probed by the observed data. In addition, we notice that a simulation length of ~ 1 Myr is long enough for the AGN to realistically span the whole ERDF, but advanced computational setup allowing simulations with $> 10^8$ points (see Section 3.2.3) will allow to probe even longer timescales.

We also note that a break position of ~30 yr is longer compared to the breaks reported in the literature for both ensemble SF or PSD, which are on the order of several years (e.g. MacLeod et al. 2012; Simm et al. 2016 C17). However, as discussed in Kozłowski (2017b), break timescales directly computed from observed SFs may be considerably underestimated due to the finite length of the AGN light curves, which is not sufficient to fully probe the variability behaviour at timescales longer than the claimed break (that is, the surveys do not satisfy $T \gg \tau_{\rm br}$). Moreover, PSD analyses almost always refer to *individual* AGN and find break frequencies which vary from source to source (e.g. Simm et al. 2016). Such short timescales breaks could therefore be "washed out" when averaged over an ensemble of sources, and could therefore not be robustly identified in the intrinsic PSD of the AGN population. Finally, we stress that the observed SF and PSD always refers to specific samples (e.g. with a magnitude limit), such that the observed variability features do not always correspond to the intrinsic ones. As an example, Fig. 13 also shows a comparison between the simulated ensemble SF and the total SF obtained from the same simulations without applying any flux cut or sample selection (i.e., the intrinsic SF). The ensemble SF obtained in Section 4.1.3 has a lower normalisation compared to the intrinsic one, suggesting that the sample selected in PTF/iPTF has an overall lower variability compared to the total sample of AGN at the same redshift and BH mass. Since the sample selection mostly relies on the AGN luminosity, including only the most luminous sources (see Section 4.1.3), this effect is consistent with works suggesting the variability to be inverse proportional to luminosity and $L/L_{\rm Edd}$ (e.g. MacLeod et al. 2010; C17; Rumbaugh et al. 2018).

| **PDF** | Broken power-law (Eq. 1) |
|---|---|
| | $\lambda^* = 0.11$ |
| | $\delta_1 = 0.47$ |
| | $\delta_2 = 2.53$ |
| | $\log(\lambda_{\rm min}) = -3.75$ |
| | $\log(\lambda_{\rm max}) = 0.5$ |
| **PSD** | Broken power-law (Eq. 3) |
| | $\alpha_{\rm low} = 1$ |
| | $\alpha_{\rm high} = 2$ |
| | $f_{\rm br} = 10^{-9}$ (~30 yr) |
| **Simulation length** | ~$1.4 \times 10^6$ yr |
| **Eddington ratio cut** | $\log(\lambda_{\rm cut}) = -1.2$ |

**Table 2.** Summary of reference simulation setup for a sample with $z \sim 1.3$, $\log(M_{\rm BH}) = 9.1$, consistent with the PTF/iPTF sample in C17.

### 4.2. *Simulating future time domain surveys*

With the advance of large, multiwavelength time domain surveys such as LSST (Ivezic et al. 2008; LSST Science Collaboration et al. 2009), TDSS (Morganson et al. 2015), ZTF (Bellm 2014), and eROSITA (Merloni et al.

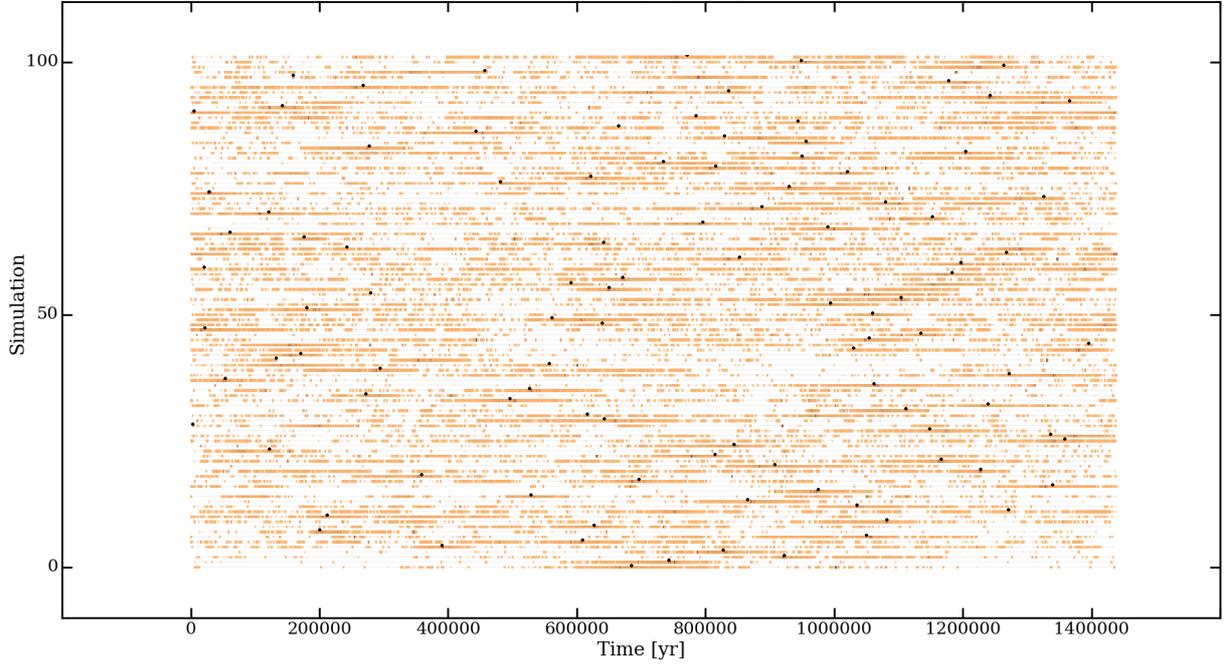

**Figure 12.** Identifying the PTF quasar sample in simulated light curves. This figure illustrates the regions above the PTF limit for 100 of the 300 reference simulations (orange; see Section 4.1.4), as well as the position of a possible PTF-observed region, selected following the procedure outlined in Section 4.1.3 (black points). Only ∼ 1.7% of the points in every light curve meet the PTF selection (mainly the SDSS+PTF flux limits). This fraction may appear to be higher in this illustration due to the limited image resolution (see, e.g., Fig. 15, top panel). Evidently, there is no preferential position within the ∼1.4 Myr covered by the simulation where the the points that could be associated with the PTF quasars sample are clustered.

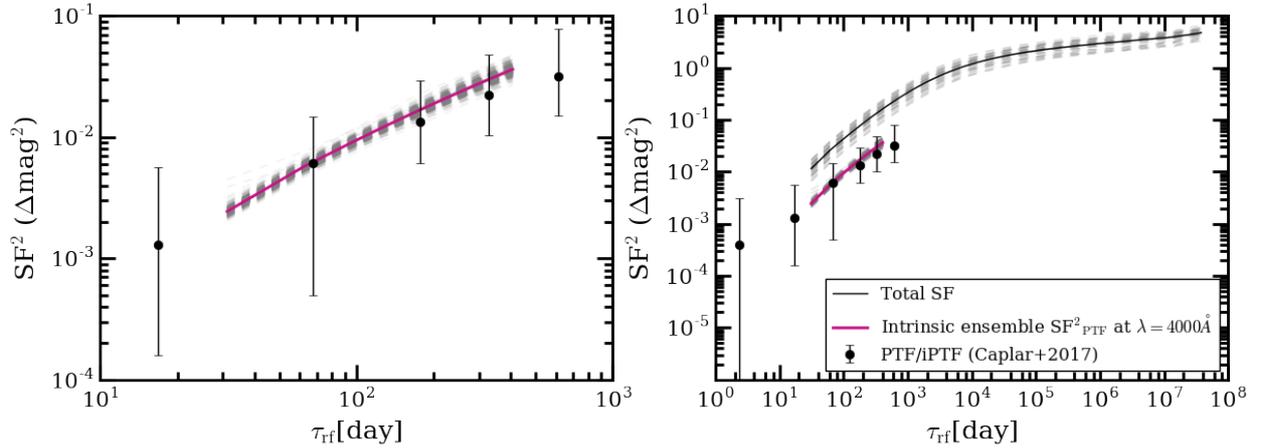

**Figure 13.** Matching the observed PTF/iPTF structure function with our reference simulation. *Left:* black points denote the ensemble PTF/iPTF SF, taken from C17; the gray lines trace 500 ensemble SF from light curves obtained with the reference simulation setup. The indigo line traces the final, representative ensemble SF for the best model, computed from the 500 intrinsic ones. *Right*: Same as left but showing also the total SF obtained directly from the best model simulations without applying any cut or sample selection (black line). We note that the ensemble SF obtained as discussed in Section 4.1.3 has a lower normalisation compared to the intrinsic one.

2012), we are entering an era of time domain astronomy which will allow us to probe the variable Universe with an unprecedented time resolution and time-span for very large statistical samples (∼millions of objects). Thanks to the high observing cadence, these surveys will allow us to probe for the first time the variability properties of *individual* AGN in a large sample, and to compare them to what was found in ensemble analyses, therefore gaining new insights on the diversity of variability behaviours among the galaxy population.

Based on current observed light curves, which show a variety of features among different AGN (e.g. Smith et al. 2018), one would expect the light curves from the new surveys to be very diverse. In our framework, this diversity arises from the fact that 1) every AGN corresponds to a different realisation of the underlying statistical process, and 2) that the observed time-span (∼ decades) is significantly shorter than the AGN life, and therefore only probe a small part of it. Indeed, we assume that the AGN variability behaviour is (statistically) the same for every AGN only "integrated" over the time needed for the AGN to span the entire ERDF, which for our reference simulation is on the order of $\sim 10^6$ yr. As a consequence, peculiar variability features lasting only for short times (i.e. rare events) will be observed only in a few galaxies, although they may be present in all of them.

To illustrate this point, Fig. 14 shows some example AGN light curves which may be expected to arise from LSST assuming the reference simulation setup defined in Section 4.1. First, we simulated a $\sim 10^6$ yr long (intrinsic frame) $L/L_{\rm Edd}$ time series based on the reference PSD and PDF (from Section 4.1) with a 9 days resolution in *observed* frame (3.9 days in intrinsic frame at $z \sim 1.3$), which is comparable to the $\sim$ weekly LSST resolution. We then converted it to bolometric luminosity and LSST $i$-band magnitude in a similar way as discussed in Section 4.1, and extracted 10 yr long "snapshots" to mimic LSST light curves. As is clearly visible in Fig. 14, every short light curve, i.e. every mock observed time series, shows a different variability behaviour. Specifically, the example shown in Fig. 14 well illustrates how the AGN can experience a variety of variability features, such as "switch on", "switch off", and a sharp, year-long "flare", as well as periods of much more subtle variability (see the different panels in Fig. 14). Naturally, not all these features will be observed in every AGN because of the limited observation length, as well as magnitude limits and host galaxy contamination (as will be discussed below). This framework therefore allows to investigate the occurrence rate of specific variability features by considering both the intrinsic AGN behaviour as well as observational limitations (e.g. the fact that we are observing every AGN only for a short time). In addition, the framework proposed in this paper will be highly valuable to constrain the true (intrinsic) AGN variability from artefacts and distortions due to the fact that we are not continuously probing the whole AGN light curve (see Section 3.2), as well as other observational biases.

One additional advantage of the new and upcoming surveys is the deeper flux limit, which will allow to detect and investigate the variability properties of fainter AGN. As an example, LSST will reach a magnitude limit $i = 24$[15], which is almost 5 magnitudes deeper than SDSS and PTF. Given these deep flux limits, it is however important to consider the possible effect of host galaxy contamination to the observed AGN light curves. To illustrate this, we again simulated a $\sim 10^6$ yr long (intrinsic frame) $L/L_{\rm Edd}$ time series as described above (Fig. 15 top, gray light curve). We then computed the $i$-band magnitude for a galaxy at $z \sim 1.3$ hosting a $\log(M_{\rm BH}) = 9.1$ BH (as in the PTF sample) by assuming the Bruzual & Charlot templates at 1 Gyr and solar metallicity (Bruzual & Charlot 2003), initial mass function (IMF) from Chabrier 2003, and the $M_{\rm BH}$-$M_{\rm bulge}$ relation from Kormendy & Ho 2013. This gives an host galaxy magnitude $i = 22.2$. The green light curve in Fig. 15 (top) corresponds to the the total light curves which considers bot AGN and galaxy contributions. The magnitude change due to host galaxy contribution is of $\sim 0.06$ mag for sources at the PTF limit, and $\sim 2$ mag at the LSST limit. Adding the galaxy contribution therefore causes an increase in total flux of $\sim 6\%$ at the PTF limit and of a factor $\sim 6\times$ at the LSST limit. Although the effect is almost negligible for PTF, it should be considered when analysing faint LSST observations.

### 4.3. *Extremely variable quasars and changing-look AGN*

In the last years, quasars exhibiting extreme variability behaviour (i.e. large magnitude changes over months-years timescales) have been found in increasingly high numbers, both serendipitously (e.g., LaMassa et al. 2015; Runnoe et al. 2016; McElroy et al. 2016; Husemann et al. 2016; Stern et al. 2018; Ross et al. 2018; Katebi et al. 2018; Trakhtenbrot et al. 2019) or through systematic searches in surveys such as the SDSS (Ruan et al. 2016; MacLeod et al. 2016; MacLeod et al. 2019; Rumbaugh et al. 2018), PanSTARRS-1 (Lawrence et al.

---

[15] This corresponds to the $5\sigma$ detection limit for a point source in a single visit (Kahn 2018).

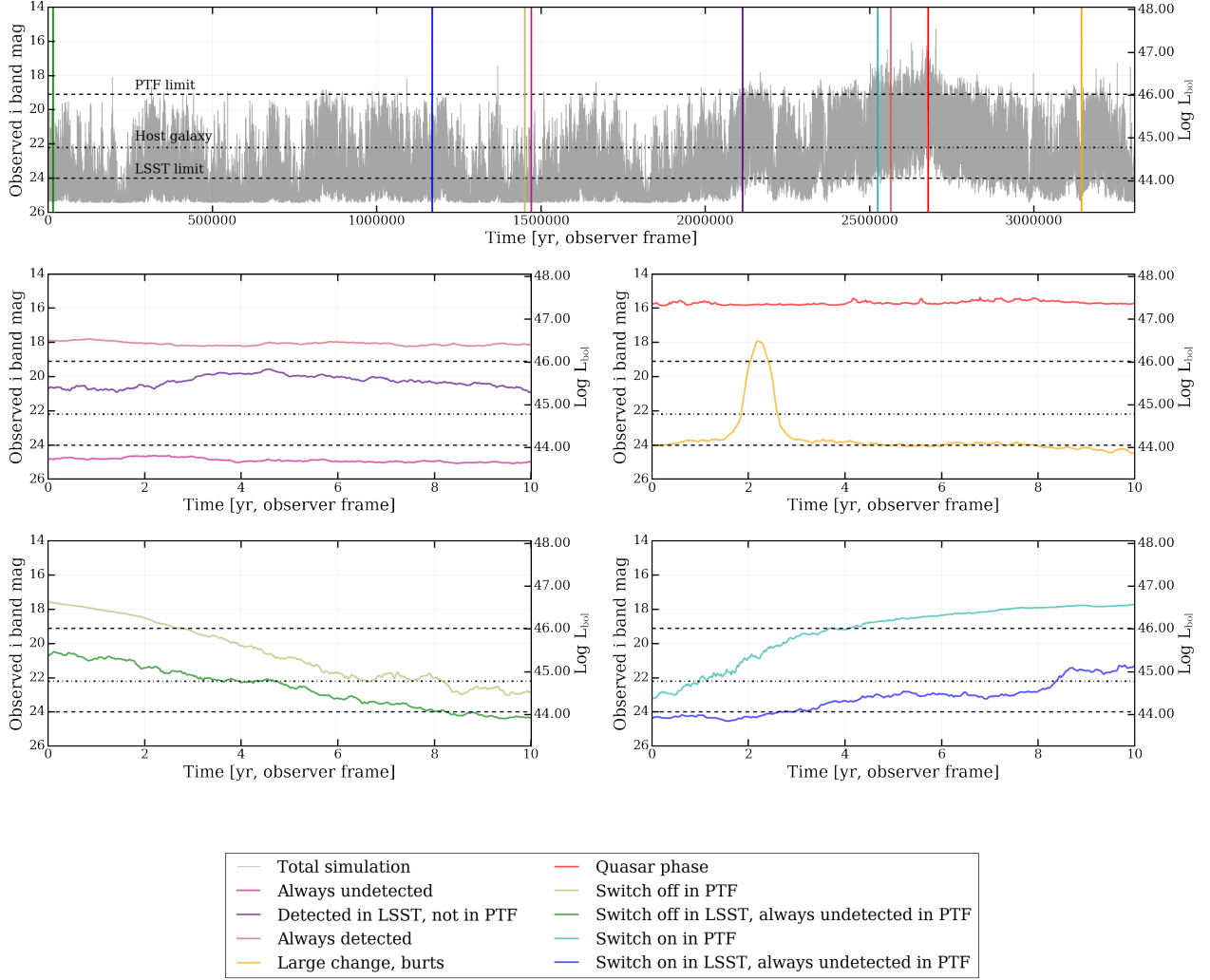

Figure 14. Simulating AGN variability as observed within the LSST. The top panel shows an example AGN light curve, simulated using the reference setup (described in Section 4.1), and a resolution of 9 days in the observed frame – comparable to the LSST cadence. The horizontal lines mark the PTF and LSST flux limits (dashed lines) and the fiducial host galaxy emission (dash-dotted line). In the panels below we show several 10 yr long "snapshots" (in the observed frame) (in the observed frame) which mimic AGN light curves that may possibly be observed within the LSST main survey. The positions within the total simulated light curve (upper panel) at which the different "snapshots" are extracted from are marked with vertical lines of the same color. This figure illustrates how observing AGN at different moments during their life can show significantly different variability features, namely: "switch on", "switch off" and "flaring" events, in addition to much more subtle AGN variability.

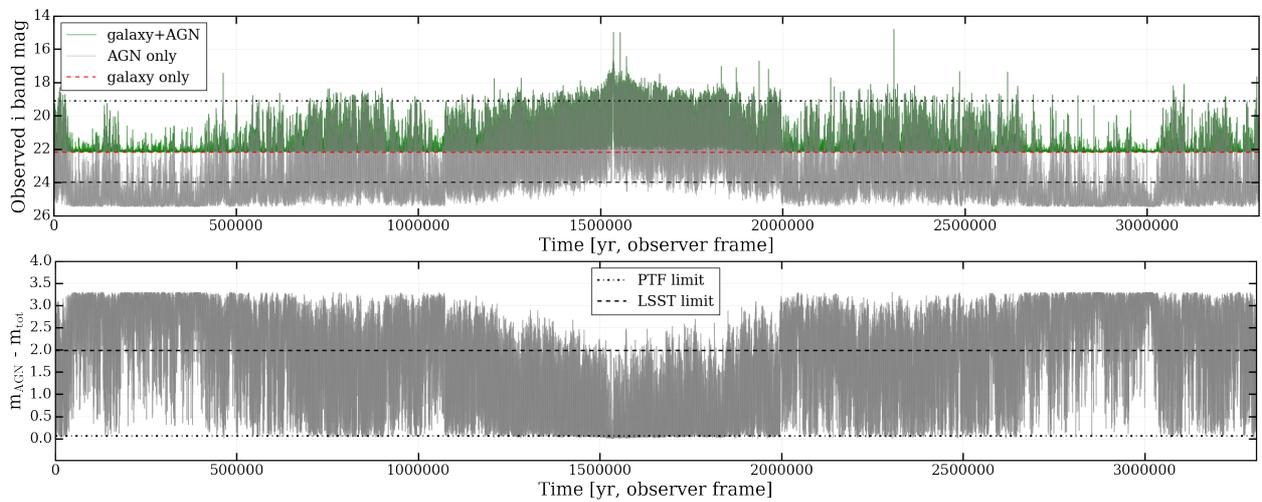

**Figure 15.** The effect of host galaxy contamination on realistic AGN light curve simulations. *Top:* in grey we show the same total AGN light curve as in the top panel of Fig. 14, but for a different realisation of the simulation (i.e., identical input parameters). This light curve corresponds solely to the AGN emission. In green we show the light curve derived when adding the contribution of the host galaxy (at $i \simeq 22$; see text for details). Since the host is fainter than the PTF flux limit, the magnitude change due to host light is only $\sim 0.06$ mag for sources above the PTF flux limit (i.e., $i \lesssim 19.1$. However, it grows to $\sim 2$ mag for the LSST limit ($i < 24$). *Bottom:* the magnitude difference due to the addition of the galaxy contribution.

2016) or CRTS (Graham et al. 2017). One example of extreme variability is observed in CL-AGN which – in addition to changes in emission line structure – show changes in luminosity of an order of magnitude or more on years-decades timescales. Future time domain surveys will allow to find and follow up in a systematic way these extreme variable objects.

Although multiple explanations for extreme variability have been put forward (e.g., Graham et al. 2017; Stern et al. 2018; Ross et al. 2018; Lawrence 2018; Rumbaugh et al. 2018), it is not yet clear if these sources are driven by physical processes that are distinct from what drives "normal" AGN variability, or if their extreme variable behaviour is simply the tail of a continuous distribution characterising the overall AGN variability phenomenon (see e.g. Sartori et al. 2018a; MacLeod et al. 2019). Our framework provides an alternative way to address these questions, by comparing the observed numbers (or rates) of variable sources with the prediction from our simulations. Indeed, our framework allows to quantify the probability of observing a given magnitude change given a model for the underlying AGN variability (i.e. the input of the simulations) and specifics of the survey (such as flux limit, cadence and length).

As an illustration, Fig. 16 shows the probability $P$ of an AGN undergoing a change in magnitude $|\Delta m| > \Delta m_{\rm ref}$ between two epochs separated by $\tau$, within our reference simulation (as specified in Section 4.1). Specifically, for a given simulated light curve, $P(|\Delta m_{\rm obs}| > \Delta m_{\rm ref})$ is defined as the ratio between the number of pairs separated by $\tau$ with $|\Delta m_{\rm obs}| > \Delta m_{\rm ref}$, and the total number of pairs separated by $\tau$. As expected, the probability of undergoing a large magnitude change drops significantly with increasing $\Delta m_{\rm ref}$, but increases with increasing $\tau^{16}$. This is a direct consequence of the assumed general PSD shape, which gives more power to low frequencies (long $\tau$).

Since the PDF assumed in this example spans the range $\log(L/L_{\rm Edd}) \in [-3.75, 0.5]$, the maximum possible $|\Delta m|$ that can be found in the simulated light curves is $|\Delta m| \sim 10.6$ mag, which corresponds to a change in luminosity of more than 4 orders of magnitude (4.24 dex). As can be clearly seen in Fig. 16, such extreme variability can be found at all timescales probed by our simulations, although the probability for such changes is extremely low. As an example, the probability for the sim-

---

[16] We caution that the trend observed at the highest probed $\Delta m_{\rm ref}$, where the probability appears to decrease for $\tau > 30$ yr, may be due to the fact that the number of corresponding pairs within the simulated light-curves is not large enough to reliably sample this low-probability regime.

ulated AGN to undergo a magnitude change $|\Delta m| > 10$ mag within $\tau = 1$ yr is $P \sim 10^{-8}$. Since such large $|\Delta m|$ over such short timescales are very unlikely to be physically explainable (see e.g. Katebi et al. 2018 and references therein for a discussion of possible timescales linked to AGN variability), this feature could arise from the fact that the assumed model (both for the PSD and for the $L/L_{\rm Edd}$ to magnitude conversion) is too simplistic, as the SED shape could change drastically (Trakhtenbrot et al. 2019). However, the probability of these features is low enough that it does not affect the main results presented in this paper, and we only acknowledge this issue to be investigated in future studies.

It is important to stress that Fig. 16 shows the probability of a simulated AGN to *undergo* a certain change in emission, which does not directly correspond to the probability of *observing* or *detecting* such a change. In fact, given the flux limit of any reasonable considered survey, combined with the contamination from the host galaxy, low $L/L_{\rm Edd}$ sources are (usually) not observed (or not identified as AGN), which limits the range of $|\Delta m|$ that may be observable.

To address this limitation, in Fig. 17 we show a similar analysis to that shown in Fig. 16, but now assuming different magnitude limits discussed in the previous sections: the PTF magnitude limit $r = 19.1$, the LSST mangitude limit $i = 24$, as well as the host galaxy contamination, at $i = 22.2$. Specifically, for a given survey depth (magnitude limit) $m_{\rm cut}$, we first substitute all the points with $m > m_{\rm cut}$ with an upper magnitude limit $m_{\rm ul} = m_{\rm cut}$, and then compute the probabilities as in the previous test, however now considering only the pairs of data points in which at least one point is above the detection limit. As expected, the probability of detecting large magnitude changes over a given timescale decreases for higher flux limits (shallower surveys). As an example, for the considered simulation the probability of detecting a magnitude change of $|\Delta m| = 2.5$ (1 dex change in luminosity) over 10 years is $P \sim 10^{-2}$ assuming the LSST depth, and $P \sim 10^{-4}$ for the PTF. We stress that the probabilities showed in Fig. 16 and Fig. 17 are not directly comparable since they are computed using different assumptions for the total number of considered pairs (total number of pairs versus number of pairs where at least one point is detected), and since Fig. 17 also includes upper limits.

If our hypothesis of an underlying common PSD+PDF set is correct, and the reference simulation defined in Section 4.1 realistically describes the AGN population, then, for sources with comparable redshifts and BH masses as the PTF quasar sample, we would expect to detect a level of variability in LSST consistent with the predictions from our simulations (after accounting for sample selection biases in a similar way to what is discussed in Section 4.1). Finding different values would however also be informative. For example, detecting a lower than expected number of highly variable quasars may indicate that the $L/L_{\rm Edd}$ changes do not directly translate to luminosity changes, but the variability is suppressed due e.g. to a luminosity dependent bolometric correction (that is, luminosity-dependent SED shape; see, e.g., Vignali et al. 2003; Marconi et al. 2004; Just et al. 2007), and to the fact that the AGN may become radiatively inefficient at low $L/L_{\rm Edd}$ (Narayan et al. 1998; Panessa et al. 2006; Ho 2009). On the other hand, an higher detection rate may be expected if additional external processes, such as microlensing (e.g. Lawrence et al. 2016), dust obscuration (e.g. Maiolino et al. 2010; Ricci et al. 2016) or supernovae also contribute to the observed luminosity changes. Other possible explanations of high variability proposed in the literature are stellar mass binary black hole mergers within the dense gas of the AGN accretion disc (e.g. McKernan et al. 2014; Bartos et al. 2017) or tidal disruption events (TDEs, e.g. Gezari et al. 2012; Arcavi et al. 2014; Hung et al. 2017; Auchettl et al. 2017). Although these events may contribute to the instantaneous emission changes seen in the concerned galaxy, their appearance is not expected to have the same probability in every galaxy (see e.g. Arcavi et al. 2014; French et al. 2016; Law-Smith et al. 2017 for the host galaxy dependence of TDEs). Therefore, such bursts of "external" origin are not expected to be present in our simulations, which are aiming at reproducing the intrinsic stochastic AGN behaviour only.

Systematic searches for CL-AGN are require large multi-epoch spectroscopic surveys, and of elaborate spectral fitting analysis to determine the changes in emission line structure (e.g. MacLeod et al. 2016). Therefore, current studies aimed at systematically searching for highly variable objects are concentrating on changes in the overall luminosity. As an example, starting from a sample of $\sim 900,000$ quasars in CRTS, Graham et al. (2017) found 51 sources with $|\Delta m| > 1$ (V-band) over $\sim 900$ days, which corresponds to a fraction of $\sim 5 \times 10^{-5} = 0.005\%$. On the other hand, Rumbaugh et al. (2018) detected of variability $|\Delta m| > 1$ (g-band) over a 15 yr period in $\sim 10\%$ of sources among a quasar sample drawn from SDSS and DES (Flaugher et al. 2015), and further estimate that this fraction would increase to $\sim 30 - 50\%$ accounting for selection effects. Furthermore, Lawrence et al. (2016) performed a joint analysis of SDSS and PanSTARRS-1 and estimated the fraction of AGN showing $|\Delta m| > 1.5$ in the g-band to lie between $10^{-5} - 10^{-3}$. A direct comparison between the results of these studies is impossible, due to the different sample selection criteria and the analysis methodology (see Rumbaugh et al. 2018 for details). Similarly, a comparison between these observational results and the predictions from simulations would require a whole analysis aimed at reproducing the sample selection and observational biases in a similar way as showed in Section 4.1 for PTF, which is beyond the scope of this paper. We stress that the proposed framework will be very valuable to test if the results from different studies can be reproduced starting from the same simple model, i.e. if the observed differences are only attributable to different sample selections and analysis methods, or if different works are finding intrinsically different classes of objects. Indeed, this framework allows to start from the intrinsic AGN variability predicted by simple models, and forward model all the observational effects which may be affecting the observed variability.

It is also important to note that, although CL-AGN usually show a change in the overall luminosity $|\Delta m| > 1$, not all the highly variable sources also present changes in the line structure. For example, MacLeod et al. (2019) report that only $10 - 50\%$ of the quasars with $|\Delta g| > 1$ in the SDSS and PanSTARRS 1 surveys also show CL-AGN behaviour in the optical spectrum. By combining upcoming photometric and spectroscopic time domain surveys such as LSST and SDSS-V it will be possible to better constrain this fraction. Together with the predictions from the simulations enabled by our framework, this will allow to investigate the relation between changes in accretion rate and changes in the broad line structure, and in general of the AGN structure, therefore gaining new insights into the interplay between accretion disc and broad line regions (e.g. Marin 2017).

## 5. SUMMARY AND CONCLUSIONS

We presented a framework designed to model AGN variability over a broad range of timescales and in different objects, based on the observed Eddington ratio distribution among the AGN population. The framework was initially proposed in Sartori et al. (2018a). In this paper we discuss in detail the fundamental assumptions of the framework; its implementation using GPU architecture; and various simulations that test and characterise both the intrinsic and numerical behaviour of our implementation. We then demonstrate several possible applications of the framework, based on the observed variability in the PTF/iPTF survey, with the aim to interpret the light curves expected to be observed with the LSST, and the prospects of discovering extremely variable AGN in up-coming time-domain surveys. An up-to-date version of the simulations code, along with examples and additional information about its usage, can be found at: https://github.com/nevencaplar/AGN-Variability-Simulations.

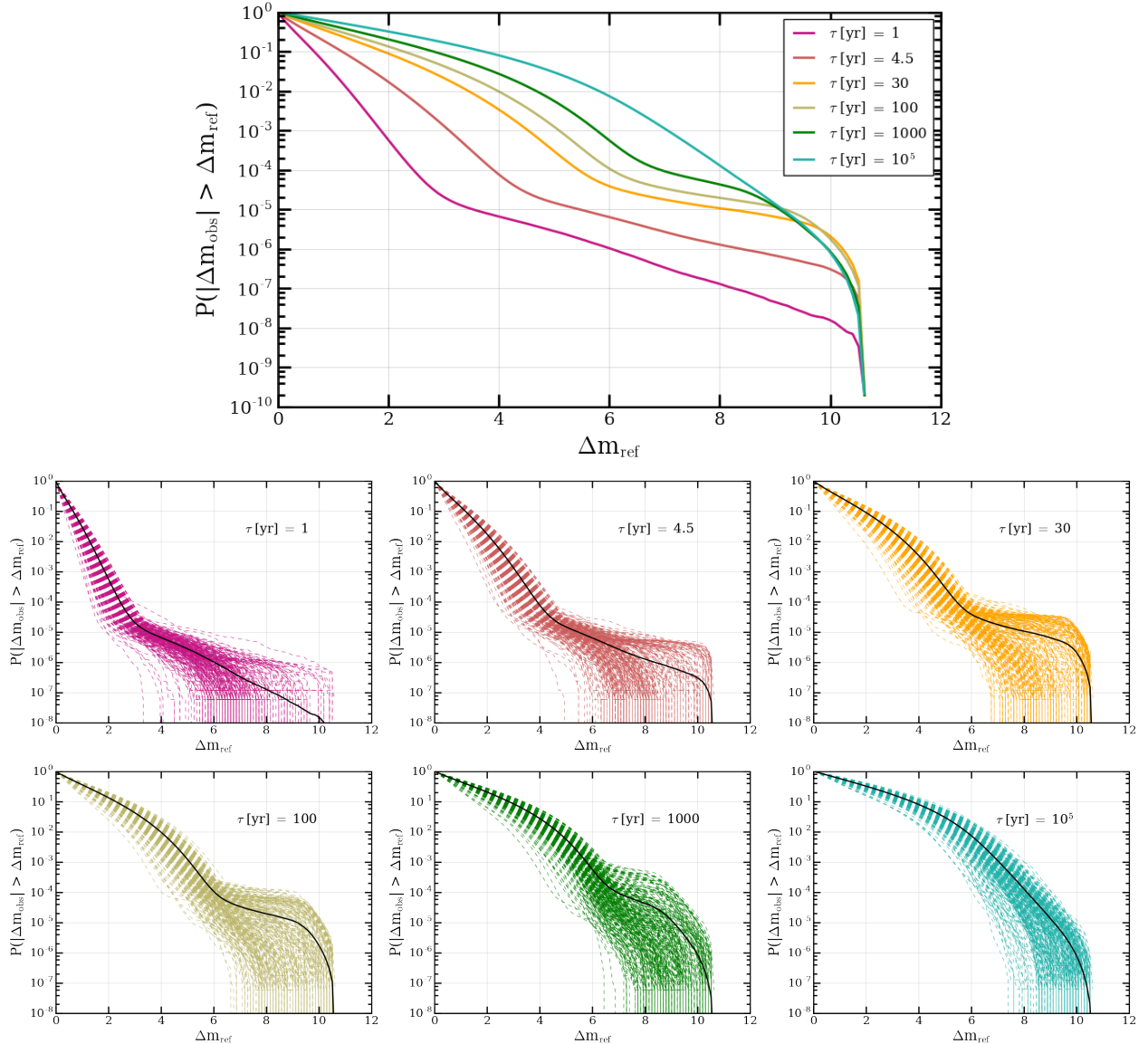

**Figure 16.** Expected distributions of AGN flux variability. We show the calculated probability of a simulated quasar, drawn from our reference simulation, to undergo a change in magnitude $|\Delta m| > \Delta m_{\rm ref}$ between two epochs separated by $\tau$ (in the intrinsic/rest frame; see Section 4.1 and Section 4.2). The small panels below show the analysis for the different $\tau$ separately, where the coloured lines are the different realisations and the black lines show the mean trends. The large panel summarises the mean trends for all the considered $\tau$. We note that at the redshift assumed for the reference simulation, $z = 1.3$, an intrinsic $\tau_{\rm int} = 4.5$ yr corresponds to an observed $\tau_{\rm obs} \sim 10$ yr, i.e. the foreseen length of the LSST main survey. Since in general the probability of the AGN undergoing a given $|\Delta m|$ decreases with decreasing $\tau$, for every possible $\tau$ probed by LSST the probability of the AGN to undergo a given $|\Delta m|$ is lower than what found for $\tau_{\rm int} = 4.5$ yr.

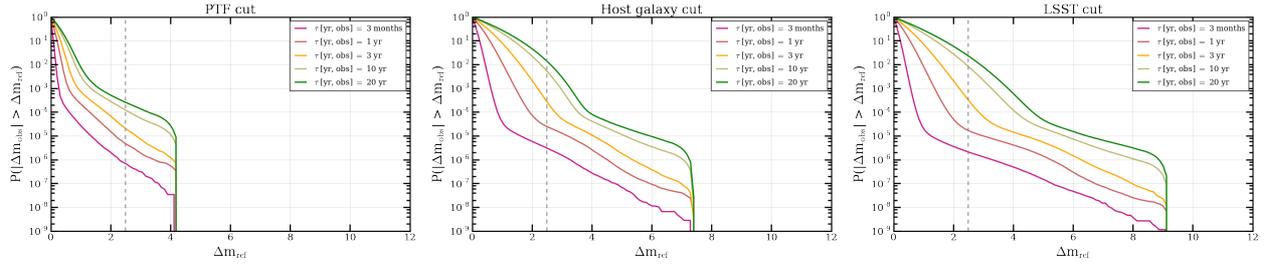

**Figure 17.** Similar to Fig. 16, but here $\tau$ is given in the observer frame, and taking into account different magnitude cuts (left to right: the PTF flux limit, the host galaxy contribution, and the LSST limit). The grey dashed lines mark a change of $\Delta m_{\rm ref} = 2.5$ (i.e., a ×10 change in flux or luminosity).

The central points of our framework can be summarised as follows see Section 2 for details):

- We propose that AGN light curves can be modeled starting from the distribution of the $L/L_{\rm Edd}$ among the galaxy population. Specifically, we assume that the distribution of $L/L_{\rm Edd}$ covered by an AGN during his lifetime is consistent with the ERDF of the galaxy population, and that the variability behaviour is described by a PSD which is similar for all the AGN.

- This implies that the variety of AGN variability features observed in different AGN arises from the fact that every AGN is a different realisation of the same underlying process (here described by the PSD and the ERDF), and that the observed period is much shorter than the AGN lifetime[17].

- The primary output of our simulations are $L/L_{\rm Edd}$ time series, which can be converted to light curves and other observables in post processing. This requires taking into account both the AGN physics (e.g. different bolometric corrections) as well as properties of the observations (e.g. flux limits and sample selection).

In order to perform the simulations we implemented the iteration algorithm from E13 to run on a GPU architecture (Section 3):

- Thanks to the CUDA environment, our GPU implementation allows to maximise the performance both in terms of length of the simulation (up to $\sim 10^8$ points) and speed ($\sim$minutes for the longest simulations).

- The code is optimised for broken power-law or log-normal PDFs and broken power-law PSDs, however different functional forms can be added.

- By construction, the PDF and the periodogram of the final light curves are consistent with the input PDF and PSD. On the other hand, the normalisation of the correspondent SF depends on the physical length (e.g. in yr) of the simulated light curve (Eq. 5).

---

[17] This can be thought of as following the ergodic hypothesis for test particles.

As an example application, we looked for a model (i.e. input set) consistent with the variability behaviour observed in the PTF/iPTF sample (C17), and used it to simulate possible light curves, of the kind that are expected to be observed with LSST (Section 4):

- After taking into account sample selection and flux limit, we found that the PTF/iPTF ensemble SF is consistent with simulations performed assuming the ERDF proposed in W17 and C17, and a broken power-law PSD characterised by a ramdom walk at low freqencies and a break at $\sim 30$ yr. In this case, the time needed for the AGN to span the entire ERDF is $\sim 1.4$ Myr (Table 2).

- We showed that the ensemble SF computed by including only light curve points above the PTF flux limit has a lower normalisation (and therefore a lower overall variability) compared to the SF computed by considering the entire simulated light curve (Fig. 13). This is consistent with observations showing that the amplitude of variability is anti-correlated with the AGN luminosity and/or Eddington ratio (e.g. MacLeod et al. 2010; C17; Rumbaugh et al. 2018).

- The simulated LSST light curves show that the same AGN can experience a variety of rather extreme variability features as "switch on", "switch off", and year-long "flares", in addition to prolonged periods of much more subtle variability (Fig. 14). Naturally, since the foreseen observations span much shorter perdios that the AGN lifetime, not all these features are observed for each AGN.

- We expect that the framework presented here would become highly valuable to understand whether any observed variability feature is consistent with the general behaviour of the AGN population (i.e. if it can be reproduced with our general model), and to constrain the intrinsic variability from artefacts due to, e.g., sample selections and observational cadence.

- Our framework provides a forward-modelling approach that can be used to compare results of different time-domain surveys, for "normal" AGN variability, extreme variability events, and, with suitable adaptations, for the variability of other astrophysical sources.

ACKNOWLEDGEMENTS

We thank the anonymous referee for helpful comments which improved the clarity of the manuscript.

LFS and KS acknowledge support from SNSF Grants PP00P2 138979 and PP00P2 166159. LFS thanks Vishal Mehta and Peter Messmer from NVIDIA Switzerland for the very useful discussions about the CUDA environment. ET acknowledges support from CONICYT-Chile grants Basal-CATA PFB-06/2007 and AFB-170002, FONDECYT Regular 1160999 and 1190818, and Anillo de ciencia y tecnologia ACT1720033.

## APPENDIX

## A. POWER SPECTRAL DENSITY AND STRUCTURE FUNCTION

A time series corresponds to a single realisation of an underlying power spectral density (PSD) and probability density function (PDF). Given a time series $x(t_k)$, $k = 1, 2, ..., N$, the statistical estimator of the underlying PSD is called periodogram $P(f_j)$ and is computed based on Fourier analysis.

Let's assume a time series $x(t_k)$, $k = 1, 2, ..., N$, with $N_{\text{steps}}$ equidistant observations with sampling period $t_k - t_{k-1} = t_{\text{bin}}$. Depending on the study, the quantity $x$ can correspond e.g. to luminosity, magnitude, counts rate or Eddington ratio. The discrete Fourier transform (DFT) of $x(t_k)$ is given by:

$$DFT(f_j) = \sum_{k=1}^{N} x(t_k) e^{2\pi i (k-1) j / N} \tag{A1}$$

where $f_j$, $j = 0, 1, ..., N-1$, are the Fourier frequencies which depend on the parity of $N_{\text{steps}}$:

$N_{\text{steps}}$ even

- $f_j^+ = j/(N t_{\text{bin}})$ for $j = 1, ..., N/2 - 1$
- $f_{N/2} = f_{\text{Nyq}} = 1/(2 t_{\text{bin}})$ for $j = N/2$
- $f_j^- = -(N-j)/(N t_{\text{bin}})$ for $j = N/2 + 1, ..., N-1$

$N_{\text{steps}}$ odd

- $f_j^+ = j/(N t_{\text{bin}})$ for $j = 1, ..., (N-1)/2$
- $f_j^- = -(N-j)/(N t_{\text{bin}})$ for $j = (N+1)/2, ..., N-1$

The periodogram $P(f_j)$ is then computed from the DFT (up to a normalisation constant depending on the convention used):

$$P(f_j) = \frac{2}{N^2} \{Re\,[DFT(f_j)]^2 + Im\,[DFT(f_j)]^2\} \tag{A2}$$

where $j = 0, 1, ..., N/2$ for even $N_{\text{steps}}$ and $j = 0, 1, ..., (N-1)/2$ for odd $N_{\text{steps}}$.

The units of the PSD, and consequently of its estimator $P(f_j)$, are [power/Hz], so that the integral of the PSD is proportional to the total power in the process:

$$\text{total power} \propto \int_0^\infty \text{PSD}(f) df \tag{A3}$$

For a given frequency $f_i$, $\text{PSD}(f_i)$ is therefore proportional to the power of variations with frequency $f_i$.

One of the most widely used PSD normalisations is the fractional rms normalisation defined by van der Klis (1997) (see also Miyamoto et al. 1991; Vaughan et al. 2003):

$$P_{\text{rms}}(f_j) = \frac{2 t_{\text{bin}}}{\mu^2 N} \{Re\,[DFT(f_j)]^2 + Im\,[DFT(f_j)]^2\} \tag{A4}$$

where $\mu$ is the mean value of the time series. With this normalisation the integral of the underlying $\text{PSD}_{\text{rms}}$ between two frequencies $f_i$ and $f_j$, $i < j$ corresponds to the contribution of the frequency window $[f_i, f_j]$ to the total rms squared variability $\sigma^2/\mu^2$. The total rms squared variability can be obtained by integrating between $f_1$ and $f_{\text{max}}$, where $f_{\text{max}} = f_{N/2} = f_{\text{Nyq}}$ for even $N_{\text{steps}}$ and $f_{\text{max}} = f_{(N-1)/2}$ for odd $N_{\text{steps}}$:

$$\frac{\sigma^2}{\mu^2} = \int_{f_1}^{f_{\text{max}}} \text{PSD}_{\text{rms}}(f) df. \tag{A5}$$

An alternative way to characterise the variability in a given time series is to construct the structure function (SF), which quantifies the amount of variability in the data for a given timescale. We stress that different definitions are present in the literature (e.g. MacLeod et al. 2012; Kozłowski 2016), so that caution has to be used when comparing SF obtained in different studies. In this work, for a time series $x(t_k)$ as defined above and a given time lag $\tau$, we adopt the following SF formulation:

$$SF(\tau)^2 = \frac{1}{P} \sum_{i,j>i} [x(t_i) - x(t_j)]^2 = \langle [x(t) - x(t+\tau)]^2 \rangle \tag{A6}$$

where $t_j - t_i = \tau$ and $P$ is the number of $\{x(t_i), x(t_j)\}$ pairs, mostly in $L/L_{\rm Edd}$ or mag units ($x(t_k)$ are the observed or simulated $L/L_{\rm Edd}$ and light curves).

In the case of a zero mean stationary time series with $f$ varying between 0 and $\infty$, a power-law PSD with slope $-\alpha$, $1 < \alpha < 3$, corresponds to a power-law SF with slope $\beta = (\alpha - 1)/2$, where $\tau = 1/f$ (e.g. Emmanoulopoulos et al. 2010). This is a direct consequence of the Wiener-Khinchin theorem. However, the analytical relationship between PSD and SF for the general case is not clear yet.